\pdfoutput=1
\documentclass[11pt,twoside,a4paper,cmspaper,final,collab]{cms-tdr}
\def\svnVersion{7d16db7}\def\svnDate{2024/09/02}\def\cmsCernNoTag{CERN-EP-2024-109}\def\cmsCernDate{\today}\def\cmsMessage{Published in the Journal of High Energy Physics as \href{http://dx.doi.org/10.1007/JHEP08(2024)215}{\doi{10.1007/JHEP08(2024)215}.}}
\begin{document}\cmsNoteHeader{EXO-22-024}

\providecommand{\cmsTable}[1]{\resizebox{\textwidth}{!}{#1}}
\newcommand{\sieie}{\ensuremath{\sigma_{i\eta i\eta}}\xspace}
\newcommand{\mgg}{\ensuremath{m_{\PGg\PGg}}\xspace}
\newcommand{\gmgm}{\ensuremath{{\PGg\PGg}}\xspace}

\newcommand{\GammaX}{\ensuremath{\Gamma_{\PX}}\xspace}
\newcommand{\mX}{\ensuremath{m_{\PX}}\xspace}

\newcommand{\mS}{\ensuremath{m_\text{S}}\xspace}
\newcommand{\mG}{\ensuremath{m_{\PXXG}}\xspace}

\newcommand{\nED}{\ensuremath{n_{\mathrm{ED}}}\xspace}
\newcommand{\MS}{\ensuremath{M_{\mathrm{S}}}\xspace}
\newcommand{\Mew}{\ensuremath{M_{\mathrm{EW}}}\xspace}
\newcommand{\Mfive}{\ensuremath{M_{5}}\xspace}

\cmsNoteHeader{EXO-22-024}
\title{Search for new physics in high-mass diphoton events from proton-proton collisions at \texorpdfstring{$\sqrt{s}=13\TeV$}{sqt(s) = 13 TeV}}

\date{\today}

\abstract{
Results are presented from a search for new physics in high-mass diphoton events from proton-proton collisions at $\sqrt{s}=13\TeV$. The data set was collected in 2016--2018 with the CMS detector at the LHC and corresponds to an integrated luminosity of 138\fbinv. Events with a diphoton invariant mass greater than 500\GeV are considered. Two different techniques are used to predict the standard model backgrounds: parametric fits to the smoothly-falling background and a first-principles calculation of the standard model diphoton spectrum at next-to-next-to-leading order in perturbative quantum chromodynamics calculations. The first technique is sensitive to resonant excesses while the second technique can identify broad differences in the invariant mass shape. The data are used to constrain the production of heavy Higgs bosons, Randall--Sundrum gravitons, the large extra dimensions model of Arkani-Hamed, Dimopoulos, and Dvali (ADD), and the continuum clockwork mechanism. No statistically significant excess is observed. The present results are the strongest limits to date on ADD extra dimensions and RS gravitons with a coupling parameter greater than 0.1.}

\hypersetup{
pdfauthor={CMS Collaboration},
pdftitle={Search for new physics in high-mass diphoton events from proton-proton collisions at sqrt(s)=13 TeV},
pdfsubject={CMS},
pdfkeywords={CMS, beyond standard model, photon production}}

\maketitle

\section{Introduction}\label{sec:intro}

While the standard model (SM) of particle physics has been a remarkably successful framework, it is widely expected to be incomplete.
A leading motivation for beyond-the-SM (BSM) physics has been the SM hierarchy problem~\cite{Weinberg:1975gm,Susskind:1978ms}, whereby quantum corrections to the Higgs boson mass must be fine-tuned to ${\sim}(\Mew/\Mpl)^2$ to keep it at the observed electroweak scale, where \Mpl and \Mew are the Planck and electroweak scales, respectively.

Large extra spatial dimensions have been proposed to resolve this problem by modifying the fundamental Planck scale.
Should the modified scale be of the same order as the electroweak scale, then little fine-tuning would be needed.
In the model proposed by Arkani-Hamed, Dimopoulos, and Dvali (ADD)~\cite{ArkaniHamed:1998rs,Antoniadis:1998ig,ArkaniHamed:1998nn}, the SM fields are constrained to the usual 3+1 spacetime dimensions, while gravity can also propagate in \nED extra, compactified spatial dimensions.
The effective strength of gravity is thus reduced by a Gauss's law reduction in the flux.

In the model proposed by Randall and Sundrum (RS)~\cite{Randall:1999ee,Randall:1999vf}, there is just one additional dimension but with a warped geometry, described by a curvature parameter $k$.
The effective strength of gravity is exponentially suppressed by the curvature over the distance in the extra dimension between the Planck ``brane'' where gravity originates and the SM brane, to which SM fields are constrained.
In both the ADD and RS scenarios, Kaluza--Klein (KK) modes of the graviton couple to the SM through the stress-energy tensor and decay into two SM particles.
In ADD extra dimensions, the KK modes are closely spaced and result in a continuum excess of diphoton events over the expected SM background.
In the RS graviton model, the KK modes are on-shell and appear as resolvable resonances in the diphoton mass spectrum.

Another proposed solution to the SM hierarchy problem is the continuum clockwork mech\-anism~\cite{Giudice:2016yja}.
In the continuum limit of the clockwork, the massless graviton is accompanied by an infinite tower of massive spin-2 graviton KK modes with a characteristic pattern of masses and couplings.
Like in the ADD extra dimension scenario, an approximately continuous distribution of KK modes results in a continuum excess~\cite{5LittleStringTheoryAtATeV,Giudiceetal}; however, in the clockwork scenario, the KK modes are all on-shell, while in the ADD case, virtual contributions to the diphoton spectrum are significant~\cite{Giudice:2003tu,Giudice:2004mg,Franceschini:2011wr}.

Finally, in addition to the above models that could address the hierarchy problem, high-mass diphoton events are also potentially sensitive to other BSM physics, such as the decays of heavy spin-0 resonances. These spin-0 resonances could arise from extended Higgs sectors~\cite{Branco:2011iw,PhysRevD.8.1226,Craig:2013hca}.

Searches for BSM physics in high-mass diphoton events from Run 2 of the CERN LHC were performed by the ATLAS and CMS experiments using proton-proton ($\Pp\Pp$) collisions at a center-of-mass energy $\sqrt{s}=13\TeV$~\cite{ATLAS:2023hbp,Sirunyan:2018wnk,Aad:2021,Aaboud:2017yyg,Khachatryan:2016yec,Aaboud:2016tru,Khachatryan:2016hje}. Prior searches have also been performed by both experiments in Run 1 of the LHC~\cite{Aad:2015mna,Khachatryan:2015qba,Aad:2012cy,Chatrchyan:2011fq,Chatrchyan:2011jx}, at $\sqrt{s}=7$ and 8\TeV, and also at the Fermilab Tevatron by the CDF~\cite{Aaltonen:2011xp,CDF:2002hrr,CDF:2010muc} and D0~\cite{Abazov:2010xh,D0:2000cve,D0:2008hxb,D0:2005srl} experiments using $\Pp\Pap$ collisions at $\sqrt{s}=1.96\TeV$.

We present the results of a search for BSM physics in high-mass diphoton events from $\Pp\Pp$ collisions at $\sqrt{s}=13\TeV$, using data collected with the CMS detector during the LHC Run~2 period of 2016--2018, corresponding to an integrated luminosity of 138\fbinv.
To search for both resonant and nonresonant deviations from the SM, we make use of two complementary background estimation techniques.
In the search for resonant excesses from both the RS and heavy Higgs boson models, we implement a technique where the diphoton spectrum is fit to a parameterized functional form, allowing for a description of the shape based exclusively on data.
In the search for nonresonant excesses that arise from the ADD and clockwork models, we use a next-to-next-to-leading order (NNLO) calculation in quantum chromodynamics of the SM diphoton background, and we estimate the background from jets being misidentified as photons using control samples in data.
In addition to using a larger sample of data, this search improves upon the previous CMS analysis~\cite{Sirunyan:2018wnk} in a variety of ways, including better photon identification and calibrations, more precise modelling of the background from jets in the case of the nonresonant analysis, and, in the case of the resonant analysis, improved statistical inference techniques.

The rest of the document is organized as follows.
The CMS detector is described in Section~\ref{sec:CMSdet}.
Event selection and reconstruction is summarized in Section~\ref{sec:event_reco_sel}.
Section~\ref{sec:signal_sim} describes the simulation of the new physics signal models in this analysis, while Section~\ref{sec:background} describes the background determination for both resonant and nonresonant excesses.
Section~\ref{sec:systematics} discusses the sources of systematic uncertainties.
We present our results in Section~\ref{sec:results} and summarize the paper in Section~\ref{sec:summary}.

Tabulated results are provided in the HEPData record for this analysis~\cite{hepdata}.

\section{The CMS Detector}\label{sec:CMSdet}

The CMS detector~\cite{CMS:2008xjf} is a multipurpose, nearly hermetic detector, designed to trigger on~\cite{CMS:2020cmk,CMS:2016ngn} and identify electrons, muons, photons, and charged and neutral hadrons~\cite{CMS:2020uim,CMS:2018rym,CMS:2014pgm}.
A global ``particle-flow'' algorithm~\cite{CMS:2017yfk} aims to reconstruct all individual particles in an event, combining information provided by the all-silicon inner tracker and by the crystal electromagnetic (ECAL) and brass-scintillator hadron (HCAL) calorimeters, operating inside a 3.8\unit{T} superconducting solenoid, with data from the gas-ionization muon detectors embedded in the flux-return yoke outside the solenoid.
The reconstructed particles are used to build tau leptons, jets, and missing transverse momentum~\cite{CMS:2018jrd,CMS:2016lmd,CMS:2019ctu}.
Within the solenoid volume are a silicon pixel and a strip tracker.

The ECAL consists of 75\,848 PbWO$_4$ crystals, each with a transverse dimension approximately matching the Moli\`ere radius of the material and 25 radiation lengths in depth.
The ECAL barrel (EB) covers the pseudorapidity range $\abs{\eta}<1.48$, and the two endcap (EE) sections cover $1.48<\abs{\eta}<3.00$.
In the EB, an energy resolution of about 1\% is achieved for unconverted or late-converting photons in the tens of \GeVns energy range.
The energy resolution of the remaining, converted barrel photons is about 1.3\% up to $\abs{\eta} = 1$, changing to about 2.5\% at $\abs{\eta} = 1.4$.
In the endcaps, the energy resolution is about 2.5\% for unconverted or late-converting photons, and between 3 and 4\% for the other ones~\cite{CMS:2020uim,CMS:2015myp}.
The diphoton mass resolution, as measured in $\PH \to \PGg\PGg$ decays, is typically in the 1--2\% range, depending on the measurement of the photon energies in the ECAL and the topology of the photons in the event~\cite{CMS:2020xrn}.

Events of interest are selected using a two-tiered trigger system.
The first level, composed of custom hardware processors, uses information from the calorimeters and muon detectors to select events at a rate of around 100\unit{kHz} within a fixed latency of about 4\mus~\cite{CMS:2020cmk}.
The second level, known as the high-level trigger, consists of a farm of processors running a version of the full event reconstruction software optimized for fast processing, and reduces the event rate to around 1\unit{kHz} before data storage~\cite{CMS:2016ngn}.
The integrated luminosities for the 2016, 2017, and 2018 data-taking years have 1.2--2.5\% individual uncertainties~\cite{CMS-LUM-17-003,CMS-PAS-LUM-17-004,CMS-PAS-LUM-18-002}, while the overall uncertainty for the 2016--2018 period is 1.6\%.
A more detailed description of the CMS detector, together with a definition of the coordinate system used and the relevant kinematic variables, can be found in Ref.~\cite{CMS:2008xjf}.

\section{Event selection}\label{sec:event_reco_sel}

Photon candidates are reconstructed from energy deposits in the ECAL, referred to as superclusters, using a clustering algorithm to associate deposits that are compatible with the expected electromagnetic shower shape along the azimuthal direction and allowing for energy recovery from electron bremsstrahlung and photon conversions.
Since the clustering algorithm does not distinguish whether the electromagnetic shower originates from a photon or an electron, it can also be applied to $\PZ \to \Pep\Pem$ events, which are then used to assess the photon energy scale and resolution, and the efficiency of the photon selection criteria.
A more detailed documentation of photon reconstruction in the CMS detector is given in Refs.~\cite{CMS:2020uim, CMS:2015myp}.

To suppress backgrounds from jets or electrons, additional selection criteria are applied to photon candidates, based on observables sensitive to the electromagnetic shower shape and any additional activity surrounding the shower.
The hadronic energy captured along the photon direction in the HCAL must be $<$5\% of the total energy of the photon.
Isolation variables are based on the total transverse momentum of particles reconstructed within a cone of size $\Delta R = \sqrt{\smash[b]{(\Delta\eta)^2+(\Delta\phi)^2}}< 0.3$ around the photon candidate, where $\phi$ is the azimuthal angle.
Two different isolation variables are constructed, summing the transverse momenta of either charged hadrons or photons.
The charged-hadron isolation is required to be less than 5.0\GeV.
The photon isolation is required to be less than 2.75 (2.0)\GeV in the EB (EE), where the isolation is corrected by the average energy density in the event to account for contributions of the photon and other interactions inside the isolation region.
The second-order moment of the log-weighted crystal energies of the photon candidate in the $\eta$ direction, \sieie, is used to measure the electromagnetic shower shape~\cite{CMS:2020uim}.
Hits in the silicon pixel and strip trackers are used to reject electrons, with a further check to ensure that photon candidates associated with electron tracks are incompatible with those resulting from photon conversions.
Finally, a selection on the shower shape variable $R_{9} > 0.8$ is imposed~\cite{CMS:2020uim}, where $R_{9}$ is defined as the energy sum of the $3\times 3$ crystals centered on the most energetic crystal in the supercluster divided by the energy of the supercluster.
The per photon reconstruction efficiency is approximately 90\% (82\%) in the EB (EE) and depends only weakly on the photon $\pt$ in the signal region.

Events are selected online by a trigger that requires at least two reconstructed photon candidates, each with transverse momentum $\pt>60\,(70)\GeV$, for the 2016 (2017--2018) data sets.
In events with more than two photon candidates, the two photons leading in \pt are chosen.
The trigger photon candidates are required to have a ratio of hadronic/electromagnetic energy of less than 0.15.
 With these selections, the trigger efficiency is close to 100\% for two photons reconstructed offline with $\pt>125\GeV$.

Two diphoton event categories are considered in this analysis: the first category is where both photons are reconstructed in the fiducial region of the EB with $\abs{\eta}<1.44$ (denoted EBEB), while the second category consists of events with one photon in the EB and the other photon in the fiducial region of the EE, $1.57 < \abs{\eta} < 2.50$ (EBEE).
Photons in the overlap region between $1.44<\abs{\eta}<1.57$ have low reconstruction efficiency and hence are not used.
The EBEE category typically provides an increase of approximately 10\% to the signal sensitivity, whereas events where both photons are in the EE are not considered because this category is dominated by SM backgrounds and has negligible sensitivity to the considered signal models.
The minimum reconstructed invariant mass of the diphoton system is required to be $\mgg > 500\GeV$.
Photon pairs must additionally satisfy $\Delta R_\gmgm > 0.45$ to be consistent with the calculation used for SM diphoton production described in Section~\ref{sec:background}.

The energy scale and resolution of photons are inferred from measurements of electrons in $\PZ \to \Pep\Pem$ events.
The energy of the photons is typically corrected by approximately 0.5\% and varies with the \pt and $\eta$ of the photon, while the additional Gaussian smearing needed in simulation to match the measured energy resolution in data varies between 0.8 and 1.5\% for photon candidates in the EB region and between 2 and 2.5\% for photon candidates in the EE region.

The interaction vertex associated with the diphoton system is selected using the algorithm described in Refs.~\cite{vertex1,vertex2}, which uses a multivariate classifier to enhance the probability that the correct vertex is assigned.
Variables that enter this classifier include the $\pt^2$ sum of the charged-particle tracks associated with the vertex, and two variables that quantify the vector and scalar balance of \pt between the diphoton system and the tracks associated with the vertex.
In addition, if either photon has an associated track that has been identified as originating from a photon conversion to an electron-positron pair, the conversion information is used in the classifier.
For signal events with diphoton invariant masses above 500\GeV, the fraction of events in which the interaction vertex is correctly assigned is approximately 90\%.

\section{Signal simulation}\label{sec:signal_sim}

Signals predictions are computed at leading order (LO) using the \PYTHIA~8.230 program~\cite{pythia82}, and the detector response is simulated with \GEANTfour~\cite{GEANT4:2002zbu}.
The signal Monte Carlo (MC) events use the NNPDF2.3~\cite{ballPartonDistributionsLHC2013} LO set of parton distribution functions (PDFs) and the CUETP8M1~\cite{khachatryanEventGeneratorTunes2016} underlying event tune for the 2016 data set and the NNPDF3.1~\cite{NNPDF:2014otw} LO set of PDFs and underlying event tune CP2~\cite{Sirunyan:2019dfx} for the 2017 and 2018 data sets.
In addition to the hard process, simulated events include the effects of additional pp interactions from within the same or nearby bunch crossings (pileup).
The simulation is reweighted such that the pileup profile matches the one observed in data, based on measurements of the luminosity and assuming that the pp inelastic cross section is 69.2~mb~\cite{CMS:2020ebo}.
The photon identification efficiency and resolution are corrected based on known differences between data and MC simulation as measured in $\PZ \to \Pep\Pem$ events.

In the resonant analysis, two values of intrinsic spin are considered: a spin-0 ($J=0$) signal, which corresponds to a heavy SM-like Higgs boson, and a spin-2 ($J=2$) signal, which corresponds to the RS graviton in the warped extra dimension model.
Three signal width hypotheses are considered for both the spin-0 and spin-2 cases, corresponding to a width that is narrower than, comparable to, and wider than the detector resolution.
In the $J=0$ case, these widths are $\GammaX/\mX = 1.4 \times 10^{-4}$, $1.4 \times 10^{-2}$, and $5.6 \times 10^{-2}$, where \GammaX is the natural width of the resonance, and \mX is the resonance mass.
In the $J=2$ case, these same intrinsic widths correspond to a relative $\widetilde{k}=0.01$, 0.1, and 0.2, where $\widetilde{k}=\sqrt{8\pi}\,k/\overline{M}_{\mathrm{Pl}}$.
Events with various mass point values are generated for each signal width in both EBEB and EBEE categories, and binned histograms in \mgg are used as the shape template.
To interpolate between generated points, a linear interpolation of the cumulative distribution functions of the \mgg shapes are used.

\begin{figure}[hbt]
    \centering
    \includegraphics[width=.60\textwidth]{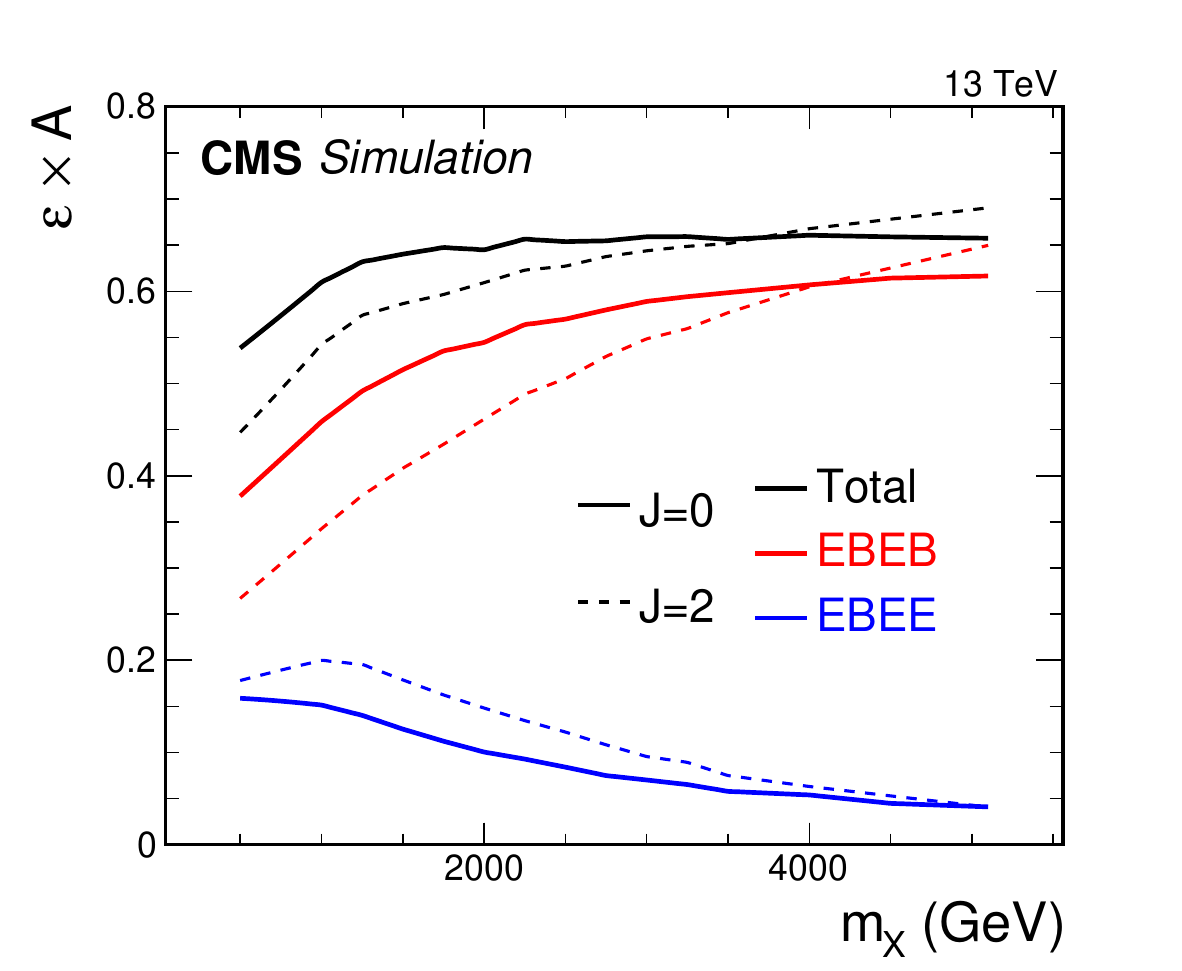}
    \caption{ The product of the acceptance ($A$) and the event selection efficiency ($\epsilon$) is shown as a function of the signal resonance mass \mX for the narrow signal width hypothesis ($\GammaX/\mX=1.4\times10^{-4}$ for $J=0$ and $\widetilde{k}=0.01$ for $J=2$). The total (black), EBEB (red), and EBEE (blue) curves are shown for spin hypotheses $J=0$ (solid) and $J=2$ (dashed).
      \label{fig:acceptance}
    }
\end{figure}

The signal normalization is determined by the product of the acceptance and the event selection efficiency.
The acceptance is estimated at the generator-level, whereas the efficiency is estimated after the full event reconstruction and selection.
The signal normalization is interpolated between each mass point with a cubic spline fit as a function of the generated mass.
In the case of the heavy Higgs boson resonance, gluon fusion is the dominant production mechanism so the off-shell production can overwhelm the on-shell signal, producing significant low-mass tails.
Thus a fiducial acceptance is quoted such that the generated diphoton invariant mass is within 20\% of the pole mass.
Figure~\ref{fig:acceptance} shows the product of the acceptance and selection efficiency of the narrow signal width hypothesis for both $J=0$ and $J=2$, as a function of \mX; the product of the acceptance and efficiency are very similar for the other width scenarios.
This figure demonstrates that the signal becomes more central as the resonance mass increases.

In the nonresonant analysis, ADD signals are generated at LO together with the SM diphoton process to account for the interference effects.
An additional SM-only sample is generated such that the difference between signal samples and the SM-only sample isolates the total effect of the ADD process.
The ADD model is parametrized by the scale variable \MS, which is the ultraviolet cutoff parameter for the virtual graviton exchange process in the Giudice--Rattazzi--Wells (GRW) convention~\cite{Giudice:1998ck}. The generated diphoton spectra are therefore truncated at the value of \MS.
Translations to the Hewett~\cite{Hewett:1998sn} and Han--Lykken--Zhang (HLZ)~\cite{Han:1998sg} conventions are performed by noting that the total cross section for the ADD signal is given by
\begin{equation}
  \sigma_{\text{total}}=\sigma_{\text{SM}}+\frac{\mathcal{F}}{\MS^4}\sigma_{\text{int}}+\frac{\mathcal{F}^2}{\MS^8}\sigma_{\text{ADD}},
\end{equation}
where $\sigma_{\text{SM}}$, $\sigma_{\text{int}}$, and $\sigma_{\text{ADD}}$ are the cross sections for the SM-only, the interference term, and the direct term, respectively. The variable $\mathcal{F}$ transforms the predictions into the different conventions,
\begin{equation}
\label{eqn:add_f_conventions}
\mathcal{F} =
\begin{cases}
1 & \text{(GRW)}, \\
\log\left(\frac{\MS^2}{\hat{s}} \right) , & \text{if } \nED = 2 \\
\frac{2}{\nED - 2} , & \text{if } \nED > 2 \\
\pm\frac{2}{\pi} & \text{(Hewett)}
\end{cases}
\biggr\}\text{(HLZ)}
\end{equation}
where $\sqrt{\hat{s}}$ is the center-of-mass energy of the colliding partons.

Predictions for the clockwork model can be obtained via a reinterpretation of the ADD signal samples.
In the clockwork model, the KK modes are all on-shell and narrow, so there is no interference effect.
Using the generated signal samples, the direct term (corresponding to $\sigma_{\text{ADD}}$) can be isolated by taking a linear combination of samples with positive interference, negative interference, and the SM diphoton process.
The direct term can then be rescaled by the expression, provided by the authors of Ref.~\cite{Giudiceetal}:
\begin{equation}
  \theta(\mgg - k) \frac{30 \Lambda^{8}_{\mathrm{T}}}{283 \pi \Mfive^{3}} \sqrt{1-\frac{k^2}{\mgg^2}} \frac{1}{\mgg^5} \left[1 + \frac{2975}{283\cdot 2^8} \left(1-\frac{k}{\mgg}\right)^9 \sqrt{\frac{\mgg}{k}} \right]^{-1},
\label{eq:rescalingfactorCW}
\end{equation}
to generate the clockwork model prediction, where $\Lambda_{\mathrm{T}}$ is equal to \MS in the GRW convention, \Mfive is the fundamental scale of the gravitational interactions, and $\theta$ is the Heaviside step function.
The clockwork spring parameter $k$ sets the inverse size of the extra dimension; phenomenologically, it controls the energy scale at which the KK modes can be excited.
Imposing perturbativity of the theory constrains $k < \Mfive$.
Because the direct term is a relatively small contribution to the overall cross section $\sigma_{\text{total}}$ at low \mgg, large statistical fluctuations in the clockwork template construction at low mass are common.
To avoid being sensitive to these fluctuations the search for the clockwork model only begins at \mgg above 1\TeV.

\section{Background determination}\label{sec:background}

The two dominant backgrounds in this analysis are the irreducible SM diphoton production and an additional reducible background from jets that mimic prompt photons (called ``misID'') in the detector.
We use two different methods to estimate the SM background.
The resonant background technique uses a functional fit to extract the background shape and is optimal for identifying narrow resonances.
The drawback of a functional fit is that it can be insensitive to shape differences that do not produce a peak, especially if the deviation lies in the tail of the distribution.
The nonresonant background technique uses a combination of higher-order MC predictions and techniques predicting the background shape from control samples in data and is best suited for identifying nonresonant deviations in the tail of the distribution.
This technique can produce large single-bin statistical fluctuations due to the misidentified-photon background, which requires a coarser binning to mitigate this effect, thus reducing the sensitivity to a narrow resonant excess.

\subsection{Resonant search background}\label{subsec:resbackground}

For the resonant signal search, parametrized functions are fit to the \mgg spectrum.
Localized excesses in the form of a peak are distinguishable from the smoothly-falling background.
Four functions of the dimensionless variable $x=\mgg/\sqrt{s}$ are given equal consideration in the statistical analysis:
\begin{equation}\begin{aligned}
f_1(x) & =  p_0  x^{p_1+p_2 \log(x)}, \\
f_2(x) & =  p_0  \re^{p_{1} x}  x^{p_{2}},  \\
f_3(x) & =  p_0 {(1+x p_{1})}^{p_{2}}, \\
f_4(x) & =  p_0  {(1+x p_{1})}^{p_{2}+p_{3} x},
\end{aligned}\end{equation}
where $p_i$ are unconstrained nuisance parameters in the hypothesis test.
During the likelihood scan of the signal strength (a numerical multiplier of the signal normalization), a discrete-profiling technique is used wherein the function that yields the maximum likelihood value is chosen in the evaluation of the likelihood.
Variable-size binning is adopted for the \mgg spectrum, so that bin widths vary from 10\GeV at low \mgg to 150\GeV at high \mgg, where the bin width is chosen to be comparable to the resolution of the reconstructed signal mass.

Tests are performed to estimate the potential bias introduced because of the background parameterization, as was done in Ref.~\cite{Sirunyan:2018wnk}.
Pseudo-experiments are generated for all background functions with different choices of assumed signal strength.
The parameterizations are considered unbiased and accurate because the mean of the measured signal strength is within $\pm$0.5 standard deviations of the injected value for all signal strengths tested.

\subsection{Nonresonant search background}\label{sec:background_nonreso}

For the nonresonant signal search, the SM diphoton production spectrum is first computed at LO with the \SHERPA~2.2.4~\cite{Sherpa} MC event generator, which includes up to three extra partons in the final state computed at tree level.
The detector response is again simulated with \GEANTfour.
A $K$~factor, defined as the generator-level ratio of the \mgg spectrum predicted by \MCFM~v8.0~\cite{Campbell:2016yrh} at NNLO to the spectrum predicted by \SHERPA, is calculated as a function of \mgg to incorporate higher-order corrections to this process.
The renormalization, factorization, and fragmentation scales in the \MCFM calculation are set to the invariant mass of the diphoton pair, \mgg.
The NNPDF30\_nnlo\_as\_0118 PDF set is used in both calculations~\cite{NNPDF:2014otw}.
Events with $\Delta R_\gmgm<0.45$ are rejected to avoid infrared divergences.
The $K$~factor values vary as a function of \mgg and range from 1.3 (1.4) to 2.0 (2.2) for EBEB (EBEE) events with $\mgg<4\TeV$.
The simulated \SHERPA events are then reweighted by the appropriate $K$~factor to produce an NNLO prediction with the full detector simulation.

The secondary background from $\PGg\text{+jet}$ and multijet events occurs when either one or two jets are misidentified as photons.
This misidentification frequently occurs when a neutral meson decays into photons at sufficiently high momentum that the photons overlap in the detector.
If there is simultaneously little hadronic energy surrounding them, these photons can look like a single, isolated photon.
A method based on control samples in data, which is similar to the one described in Ref.~\cite{Sirunyan:2018wnk}, is used to estimate the rate in each \mgg bin.

This method computes a transfer ratio in two reference samples, a jet-triggered data set and a double-muon-triggered data set.
The two samples are enriched in jets produced through the strong interaction and in association with a $\PZ$ boson, respectively, and have different quark-gluon compositions that are used to bound the uncertainty in the misID rate.
The numerator of this ratio is the number of jets that are misidentified as photons.
The denominator is the number of jets that are reconstructed as `photon-like' objects, where photon-like means that the object passes a less-strict version of the photon identification criteria and also fails at least one of the isolation or shower shape requirements.
The contamination of the numerator of the transfer ratio by prompt photons is estimated statistically by fitting the shower shape variable \sieie to two templates.
The contamination of the denominator by prompt photons is negligibly small.
The transfer ratio can thus be used to estimate the number of jets misidentified as photons in a signal sample from the number of photon-like jets in that same sample.

This transfer ratio is measured as a function of the photon object \pt, the number of reconstructed primary vertices, and in three $\abs{\eta}$ regions: the EB and two ranges in the EE, corresponding to $1.57 < \abs{\eta} < 2.03$ and $2.03 < \abs{\eta} < 2.50$.
The final estimate of the misID background sources is calculated by measuring the number of events that pass the final event selection but with either one or two photon-like objects replacing one or two photons and applying the transfer ratio to those events.
This is performed for each \mgg bin separately for the EBEB and EBEE categories.

\section{Systematic uncertainties}\label{sec:systematics}

Sources of systematic uncertainty held in common by the two background treatments include a 1.6\% uncertainty in the signal normalization, stemming from the uncertainty in the total integrated luminosity.
Uncertainties in the signal shape come from the photon identification efficiency, the photon energy scale, and the photon energy resolution.
In the range $1<\mgg<3\TeV$, the maximum effect of these uncertainties in the rate per bin are approximately 7--15\% for the photon identification efficiency and 2--4\% for the photon energy scale.
The uncertainty in the energy resolution, including the vertex assignment accuracy, is $<$1\%, and its effect is negligible for the nonresonant analysis.
The uncertainty due to the pileup is assigned by changing the total inelastic cross section by $\pm$4.6\% in the pileup reweighting procedure~\cite{Sirunyan:2018il}.

In the resonant analysis, there is an additional 6\% uncertainty in the signal normalization in order to account for the variation in the kinematic acceptance of the signal that comes from the use of alternative PDF sets.
The resonant analysis also treats the background shape parameters described in Section~\ref{subsec:resbackground} as unconstrained nuisance parameters.

In the nonresonant analysis, there are several sources of uncertainty associated with the backgrounds.
The diphoton background has a 1.6\% uncertainty in its normalization due to the uncertainty associated with the integrated luminosity, which is assumed to be fully correlated with the luminosity uncertainty in the signal.
It also has an additional 5\% normalization uncertainty from potential higher-order corrections beyond the NNLO calculation.
Uncertainties that modify the background shape arise from both theoretical and experimental sources.
The PDFs are treated as 50 independent eigenvectors with uncorrelated uncertainties.
Uncertainties in the $K$~factors arise from changing the renormalization, factorization, and fragmentation scales simultaneously between $\mgg/2$ and $2\mgg$.
Uncertainties due to pileup, as well as the photon efficiency and energy scale, are applied to background in the same way as for signal events.
For the misID photon background contribution, an uncertainty in the shape as well as an overall 30\% uncertainty in the normalization is assigned to cover variations observed in the rates measured in the different data sets and from closure studies based on simulation samples.

The systematic uncertainties are treated as nuisance parameters in the calculation of the final results.
They are assumed to have a log-normal shape profile, unless otherwise stated.

\section{Results}\label{sec:results}

\begin{figure}[hbt]
\centering
    \includegraphics[width=.45\textwidth]{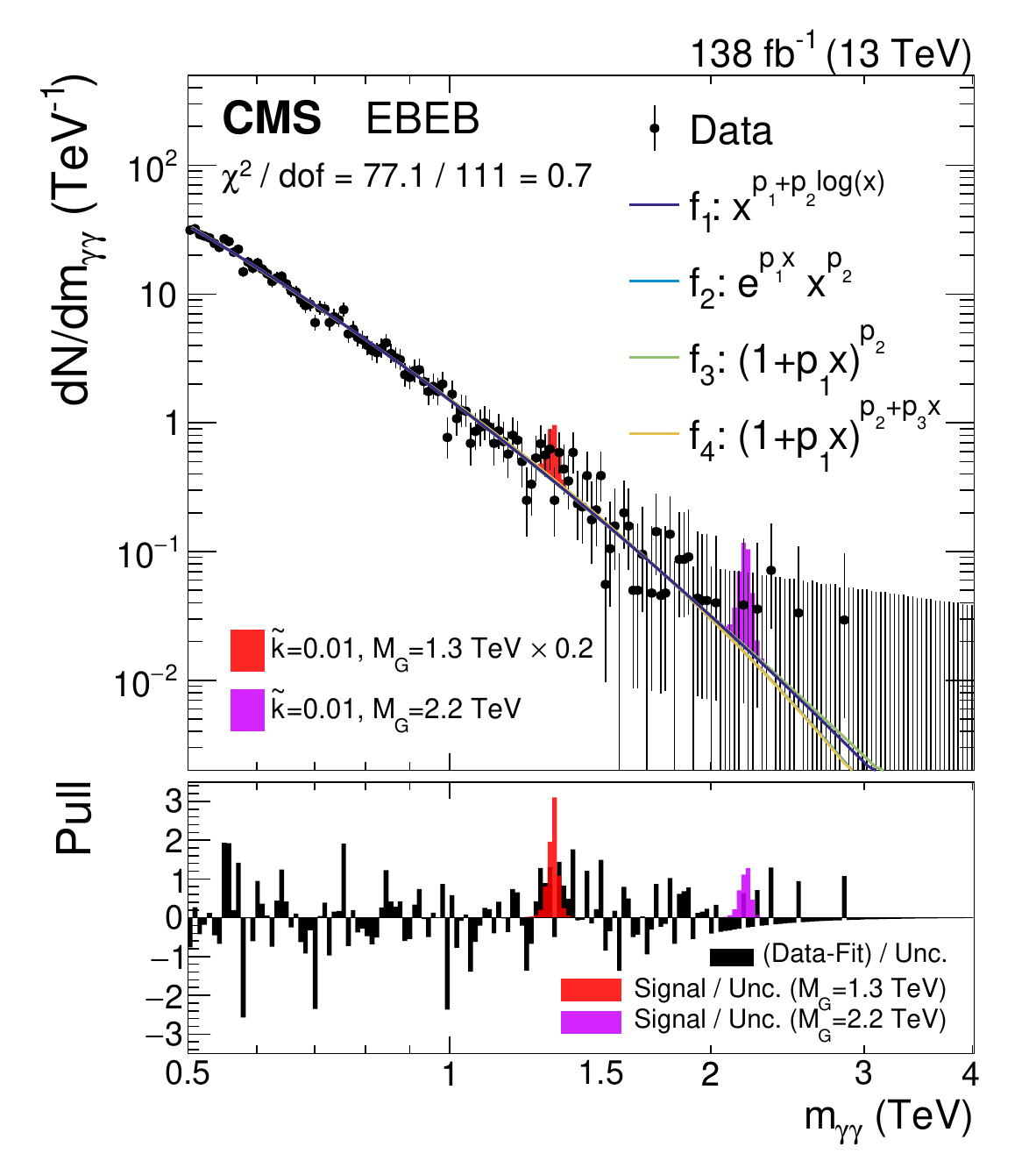}
    \includegraphics[width=.45\textwidth]{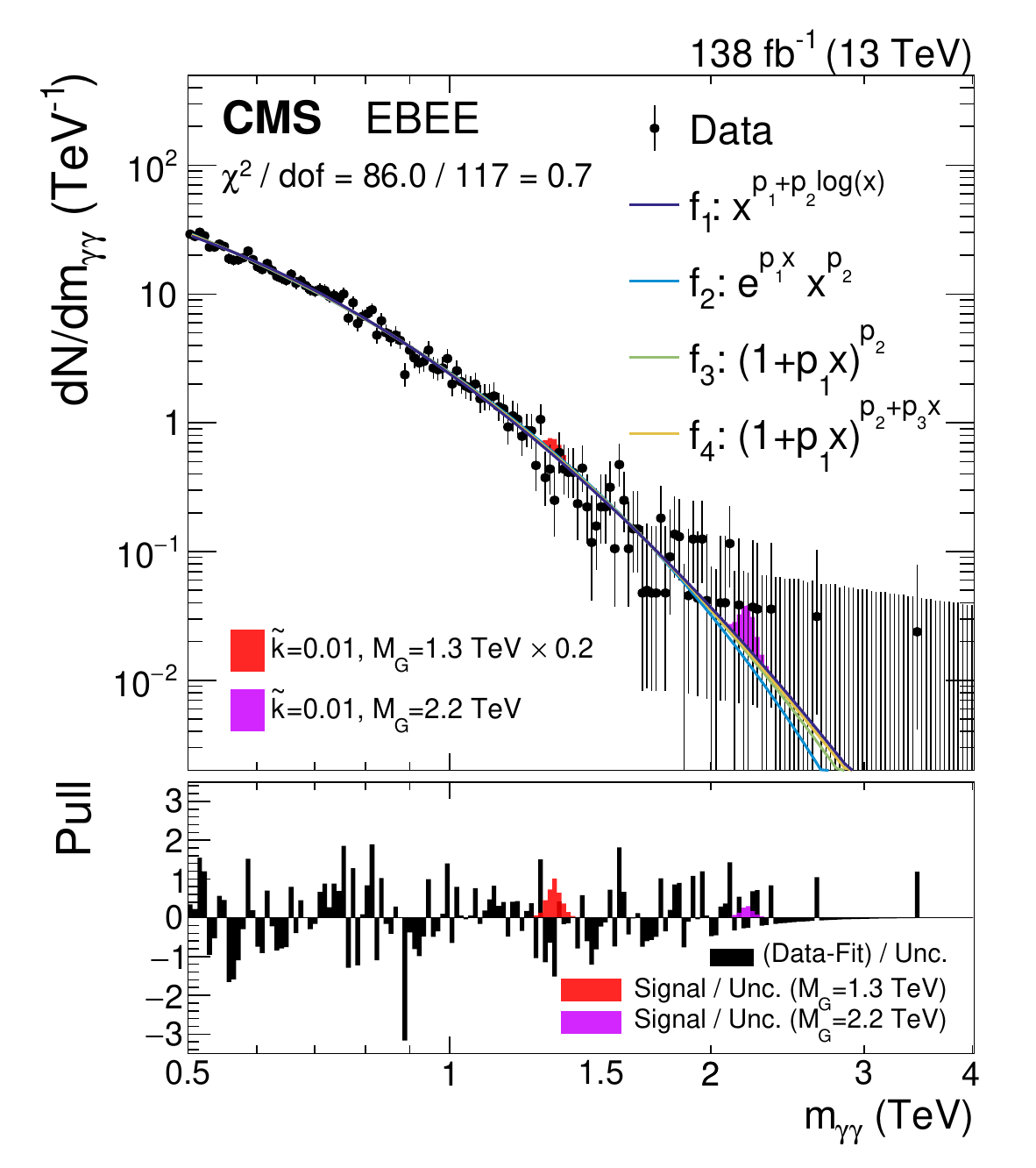}
\caption{Observed diphoton invariant mass spectra for the EBEB (left) and EBEE (right) categories for the 2016--2018 data are shown. Also shown are the results of a likelihood fit to the background-only hypothesis. The black, blue, green, and yellow lines indicate the result of the fit functions $f_1$, $f_2$, $f_3$, and $f_4$, respectively. The predicted excesses from narrow RS gravitons at masses 1.3 and 2.2\TeV are shown based on the theoretical LO cross sections, with the 1.3\TeV signal scaled by an additional factor of 0.2. The lower panels show the difference between the data and the $f_1$ fit, divided by the statistical uncertainty in the data points. The indicated $\chi^2$ in the plot is also given with respect to the $f_1$ fit.
\label{fig:paramfit}
}
\end{figure}

The results of the resonant search are obtained assuming two kinds of signals, RS gravitons and heavy Higgs bosons.
A composite hypothesis test is conducted to compare the agreement of observed data with the null and alternative hypotheses, which are the background-only and the signal-plus-background hypotheses, respectively.
A simultaneous fit is performed to the \mgg spectrum in 6 event categories, EBEB and EBEE, separately for each of the three years of data taking.
The test statistics are based on the profile likelihood ratio:
\begin{equation}
q(\mu) = -\log \frac{ L(\mu S + B | \hat\theta_{\mu} ) } {L (\hat\mu S + B | \hat\theta)},
\end{equation}
where $S$ and $B$ are the probability distribution functions for the signal and SM background processes, respectively, $\mu$ is the signal strength parameter, and $\theta$ are the nuisance parameters of the model.
The ``hat'' notation indicates the best-fit value of the parameters.
The result of the fits, combining the 2016, 2017, and 2018 data sets, using bin sizes chosen for the narrow width hypothesis are shown in Fig.~\ref{fig:paramfit}.

\begin{figure}[hbt]
    \centering
    \includegraphics[width=.45\textwidth]{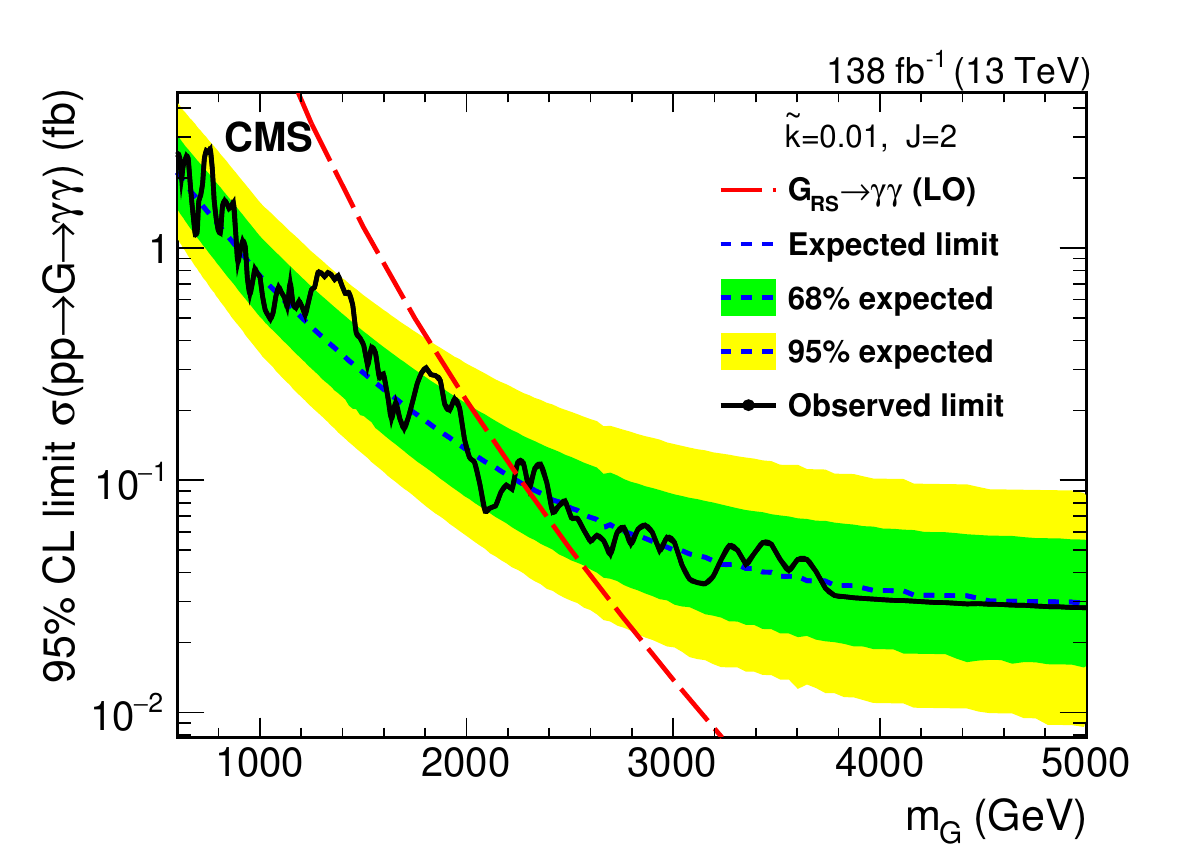}
    \includegraphics[width=.45\textwidth]{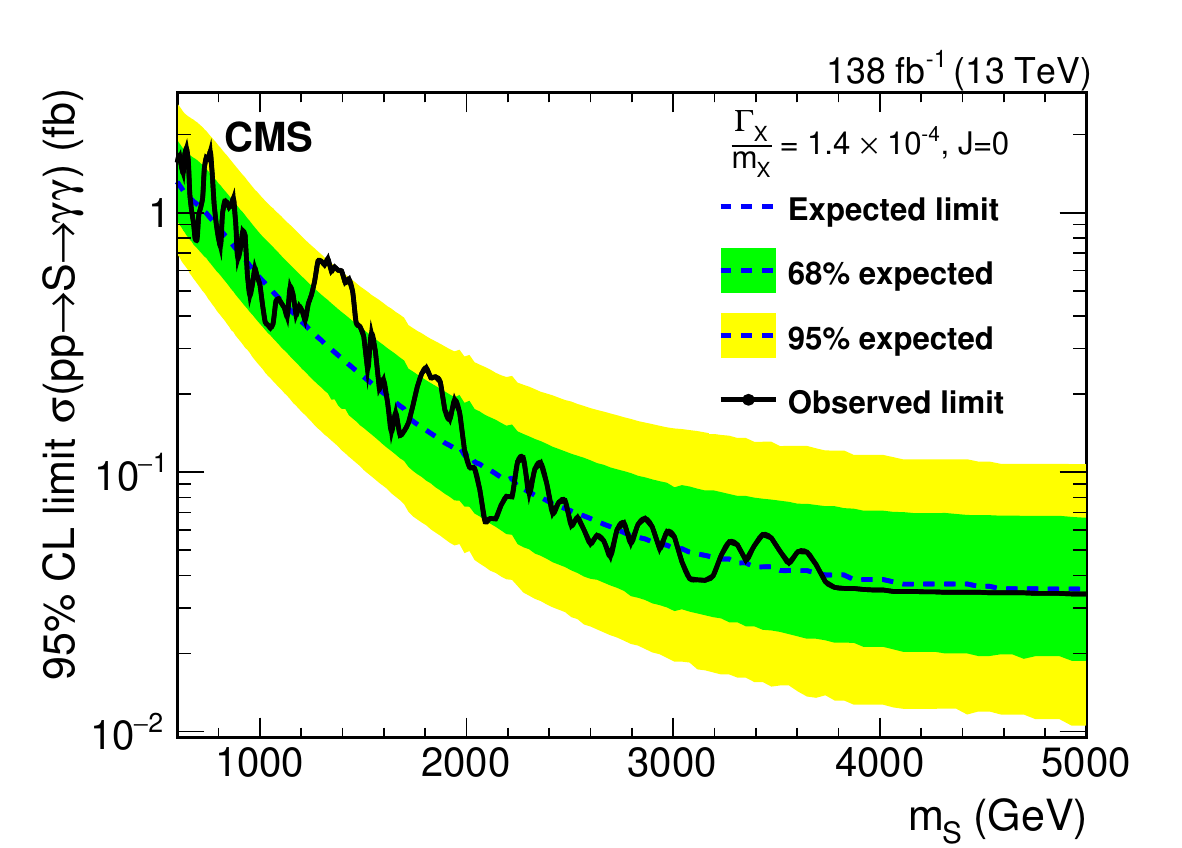} \\
    \includegraphics[width=.45\textwidth]{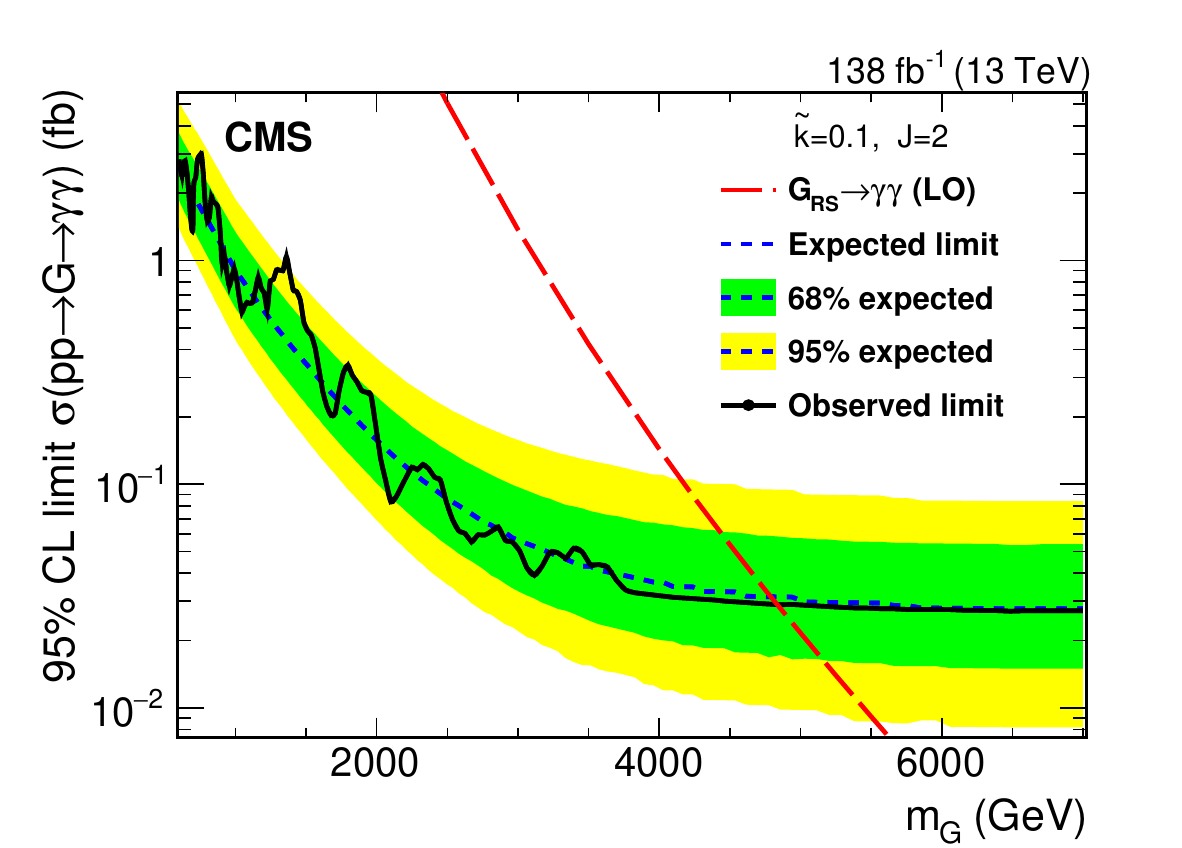}
    \includegraphics[width=.45\textwidth]{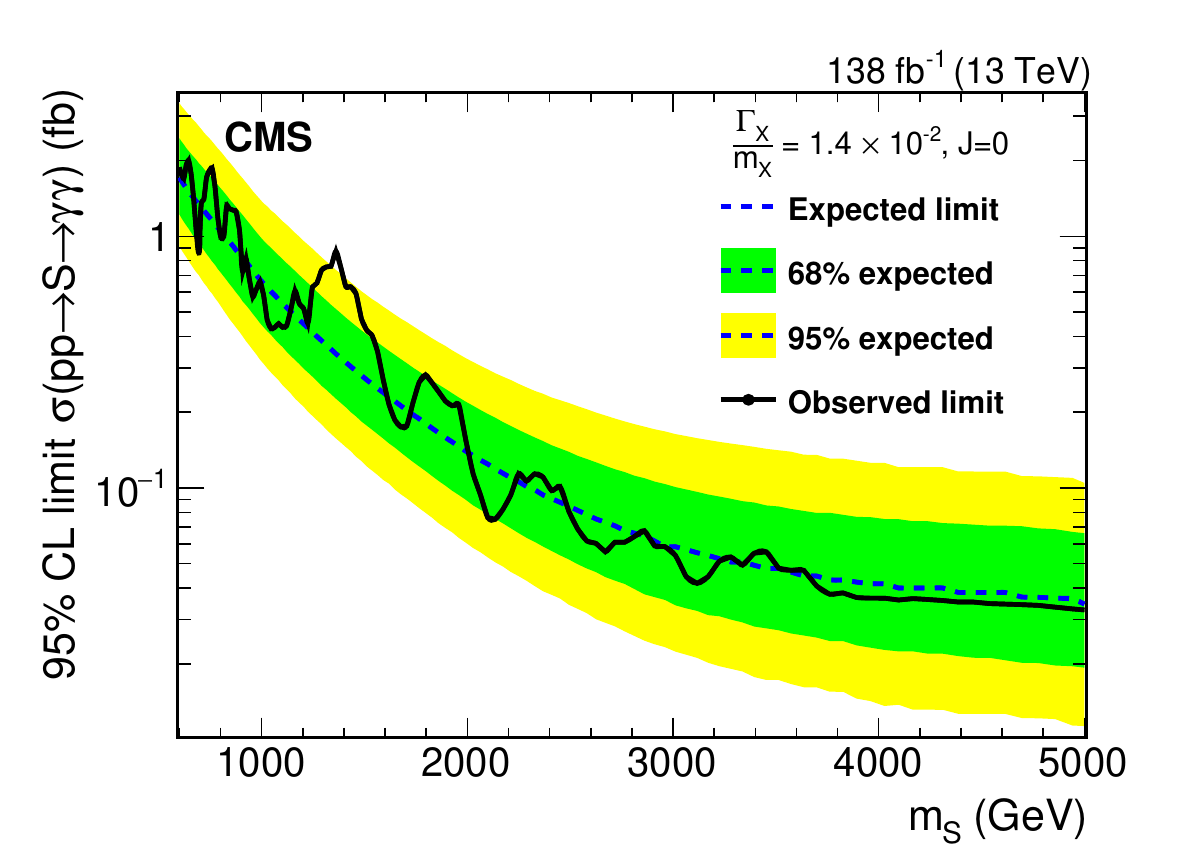} \\
    \includegraphics[width=.45\textwidth]{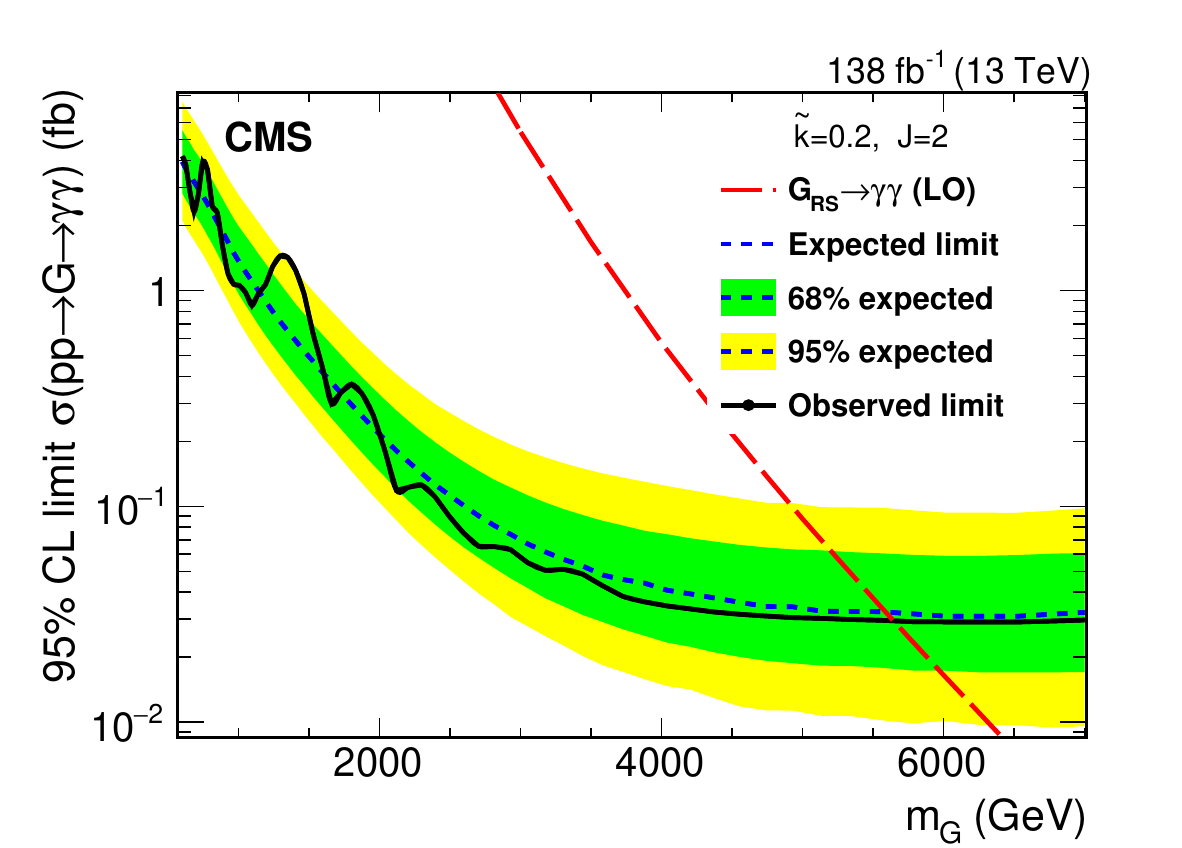}
    \includegraphics[width=.45\textwidth]{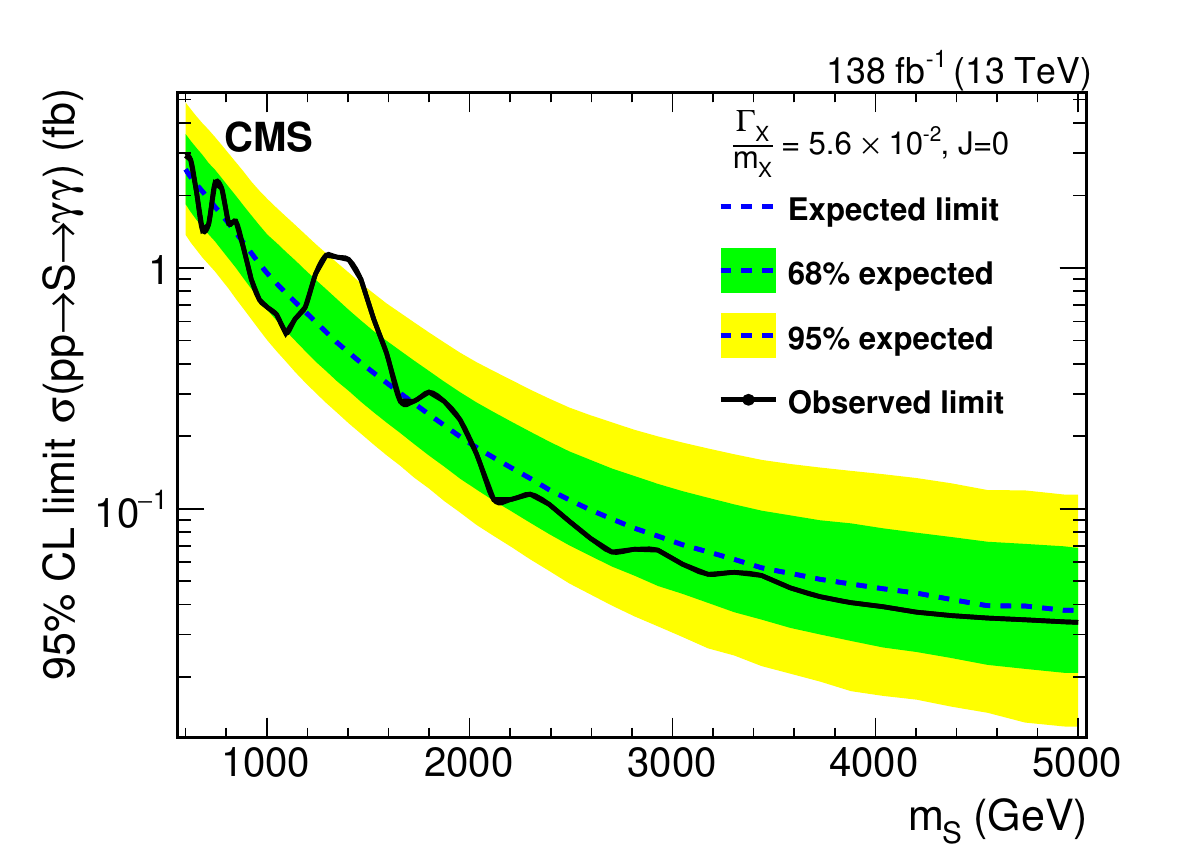} \\
    \caption{Expected and observed 95\% \CL upper limits on the product of the production cross section and branching fraction as a function of the RS graviton mass \mG (left) and heavy Higgs boson mass \mS (right) for the full Run 2 data set.
    The dotted red line is the LO theoretical cross section for the RS graviton.
    The rows correspond to different resonance widths.
    Expected 68\% and 95\% limit bands are shown in green and yellow, respectively.
    \label{fig:limits_fullRun2}
    }
\end{figure}

\begin{figure}[!htb]
    \centering
    \includegraphics[width=.45\textwidth]{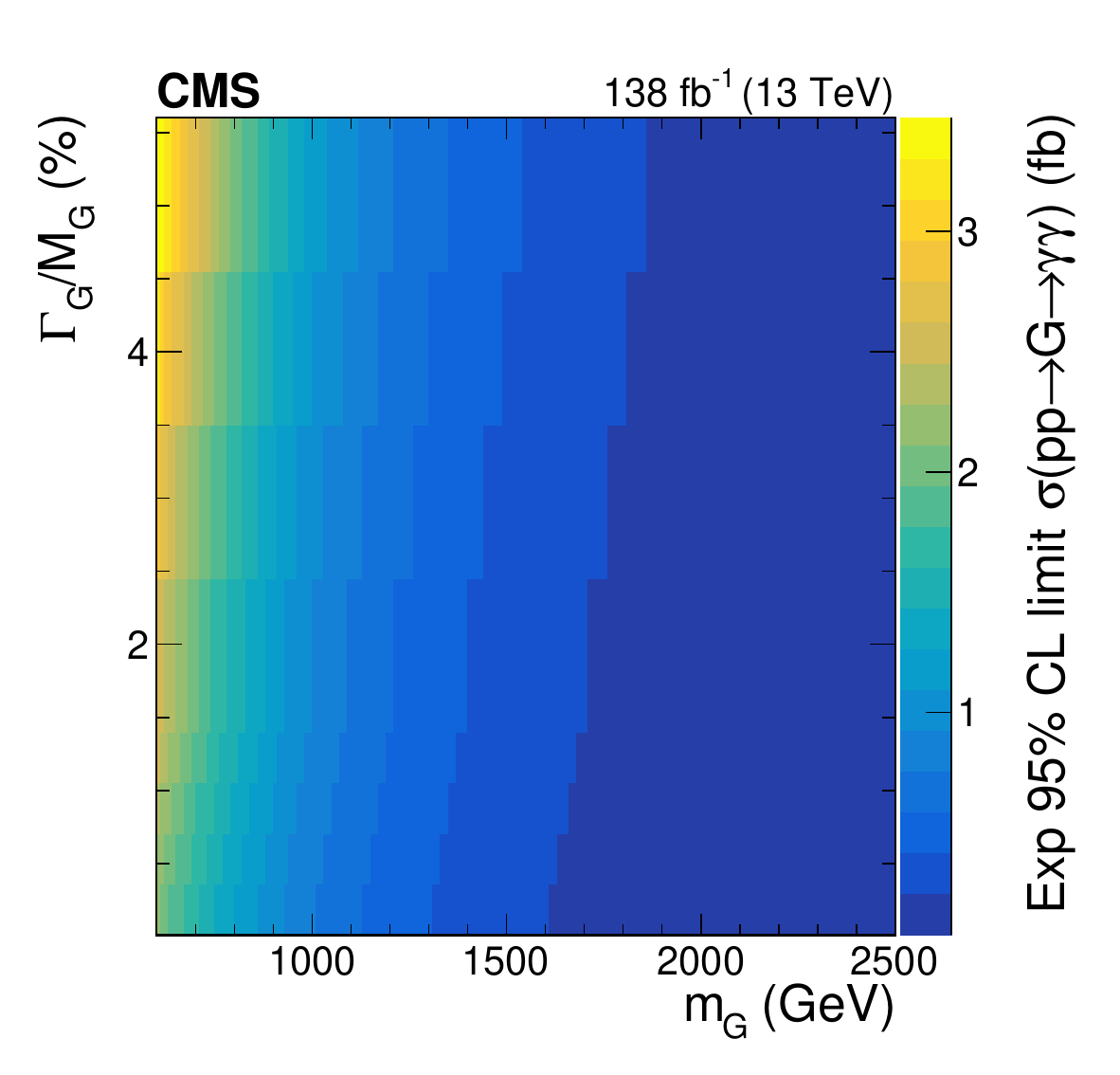}
    \includegraphics[width=.45\textwidth]{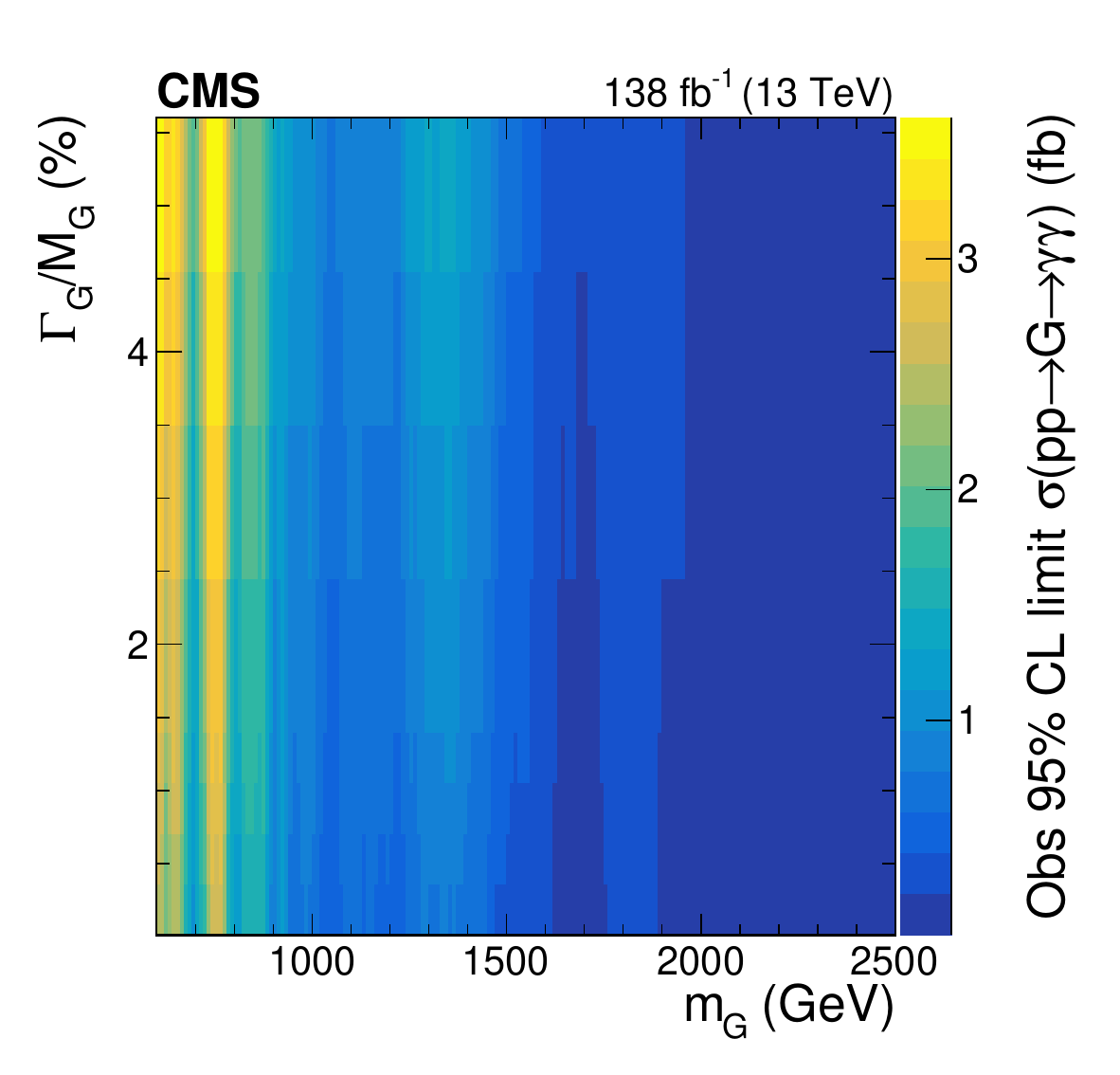} \\
    \includegraphics[width=.45\textwidth]{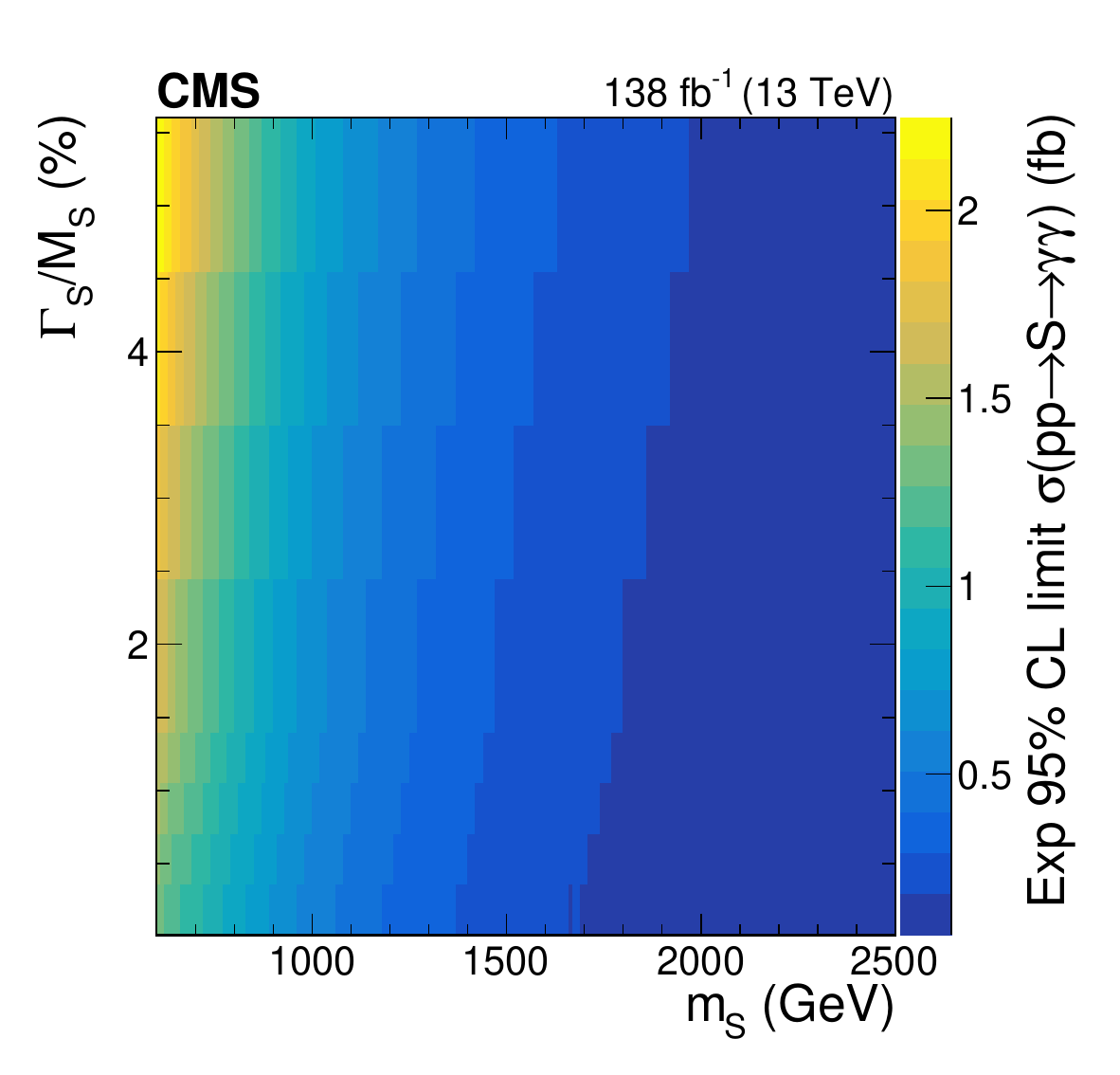}
    \includegraphics[width=.45\textwidth]{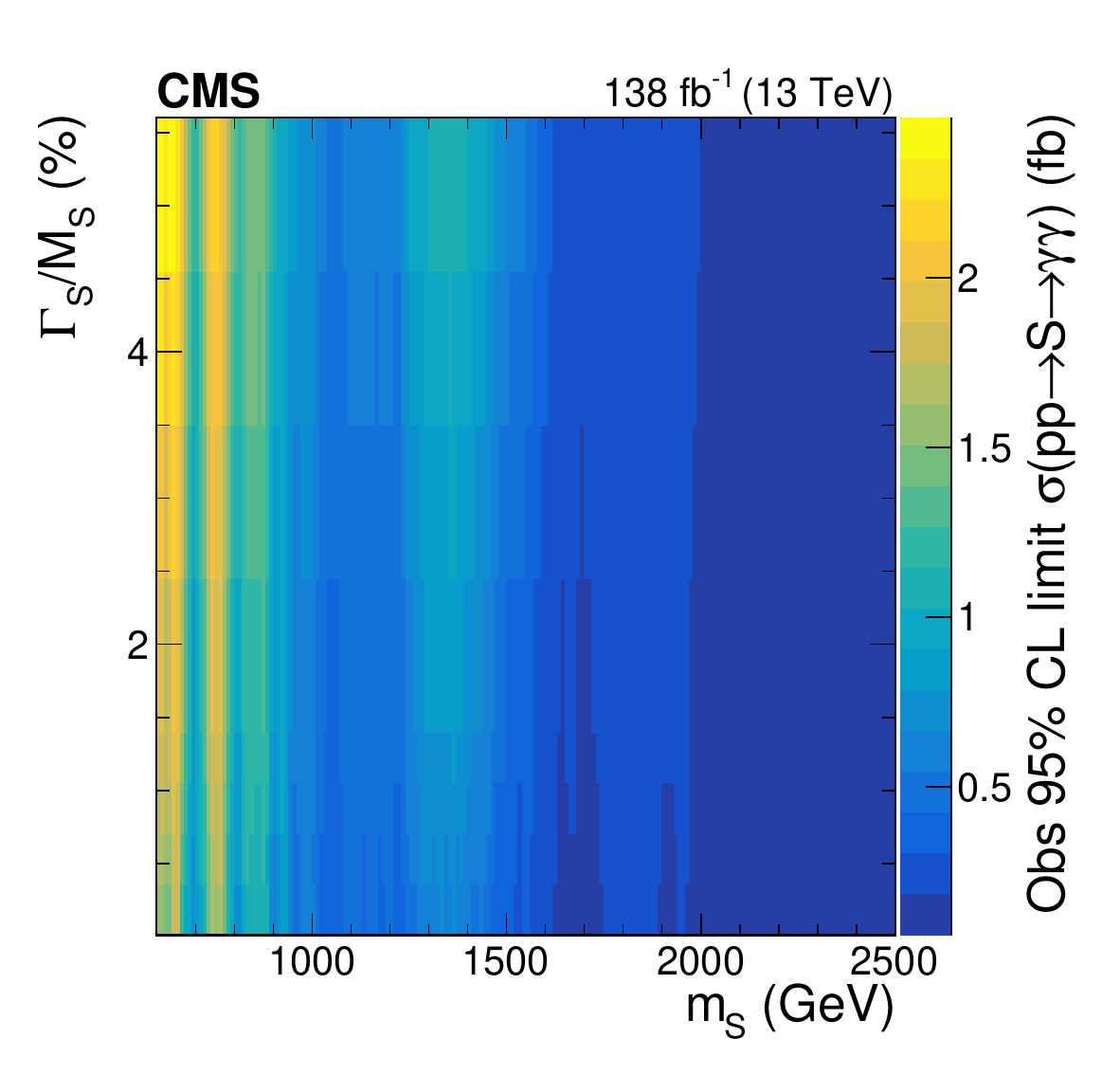} \\
\caption{ Expected (left) and observed (right) 95\% \CL upper limits on the product of the cross section and branching fraction as a function of the RS graviton mass \mG (upper) and heavy Higgs boson mass \mS (lower) versus the resonance width for the 2016--2018 data.\label{fig:limits_2D}}
\end{figure}

No statistically significant excess is observed in the data.
To set the upper limit on the signal cross section, the modified frequentist method, commonly known as \CLs, is used according to the prescriptions of Ref.~\cite{ATLAS:1379837}, in conjunction with the use of asymptotic formulas~\cite{Cowan:2010js}.
The 95\% confidence level (\CL) upper limits on the production cross sections are reported in Fig.~\ref{fig:limits_fullRun2}, as a function of the resonance mass.
Figure~\ref{fig:limits_2D} reports the same expected and observed upper limits but as a function of the resonance mass and width.
The largest excess over the SM expectation is found at 1320\GeV with a width of $\GammaX/\mX=5.6\times10^{-2}$, and the local significance is 2.6 standard deviations.
The global significance, computed by considering resonance mass hypotheses from 0.6--2.5\TeV and widths from $1.4\times10^{-4}$ to $5.6\times 10^{-2}$, is 0.8 standard deviations.

For the nonresonant analysis, constant bin widths of 100\GeV are used.
Similar to the resonant analysis, a simultaneous fit to the \mgg spectrum in six event categories, EBEB and EBEE for each of the three years of data collection, is performed.
The \mgg spectrum begins at 500\GeV for the ADD interpretation, while this threshold is raised to 1\TeV in the clockwork mechanism analysis to avoid statistical fluctuations in the template construction described in Section~\ref{sec:signal_sim}.
A Bayesian statistical approach was taken to constrain the signal parameters with a flat prior on the signal strength, which is constrained to be positive, and a Markov chain MC method~\cite{Metropolis:1953am} is used to marginalize the nuisance parameters.

\begin{figure}[!htbp]
\centering
\includegraphics[width=0.48\textwidth]{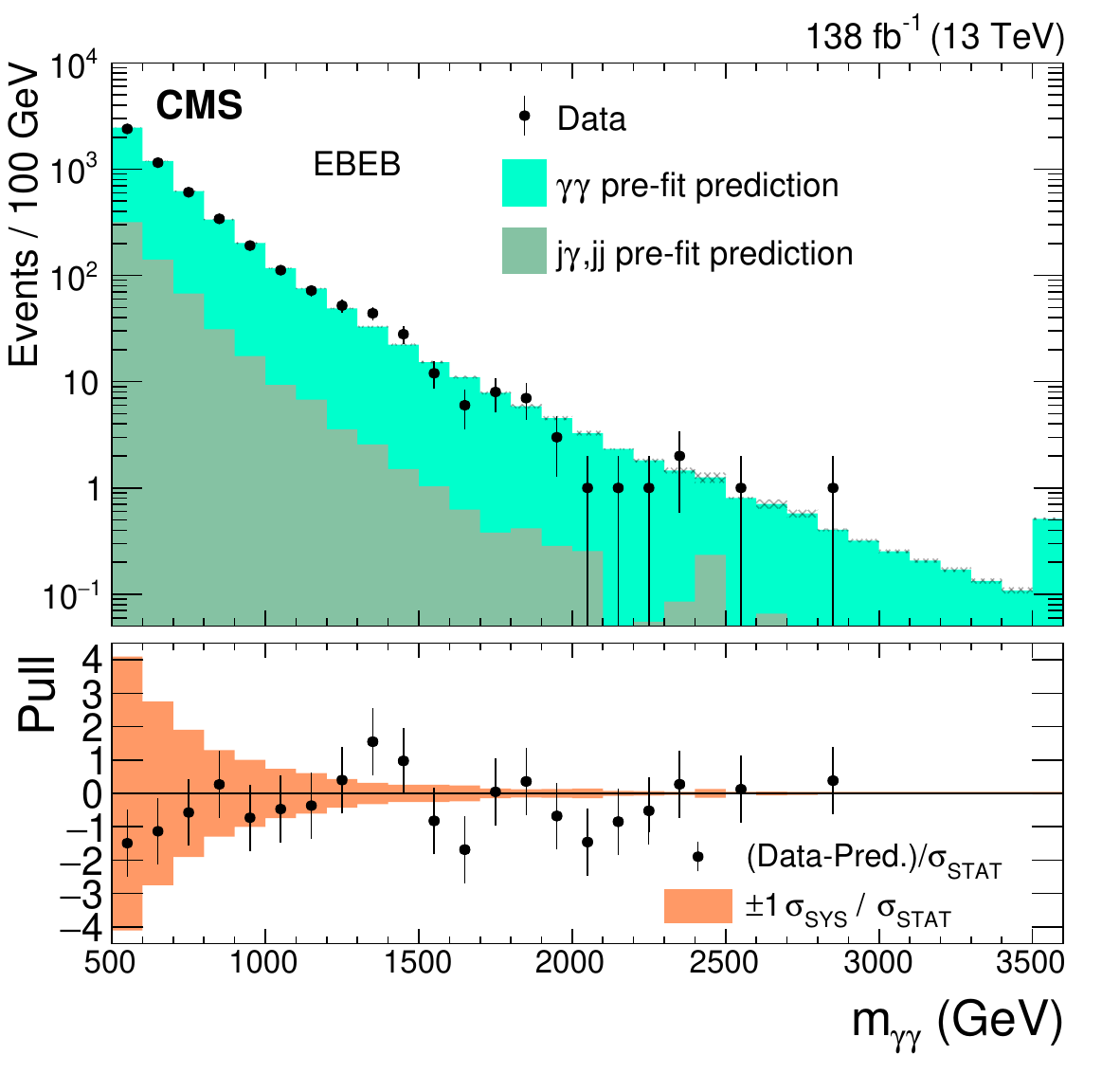}
\includegraphics[width=0.48\textwidth]{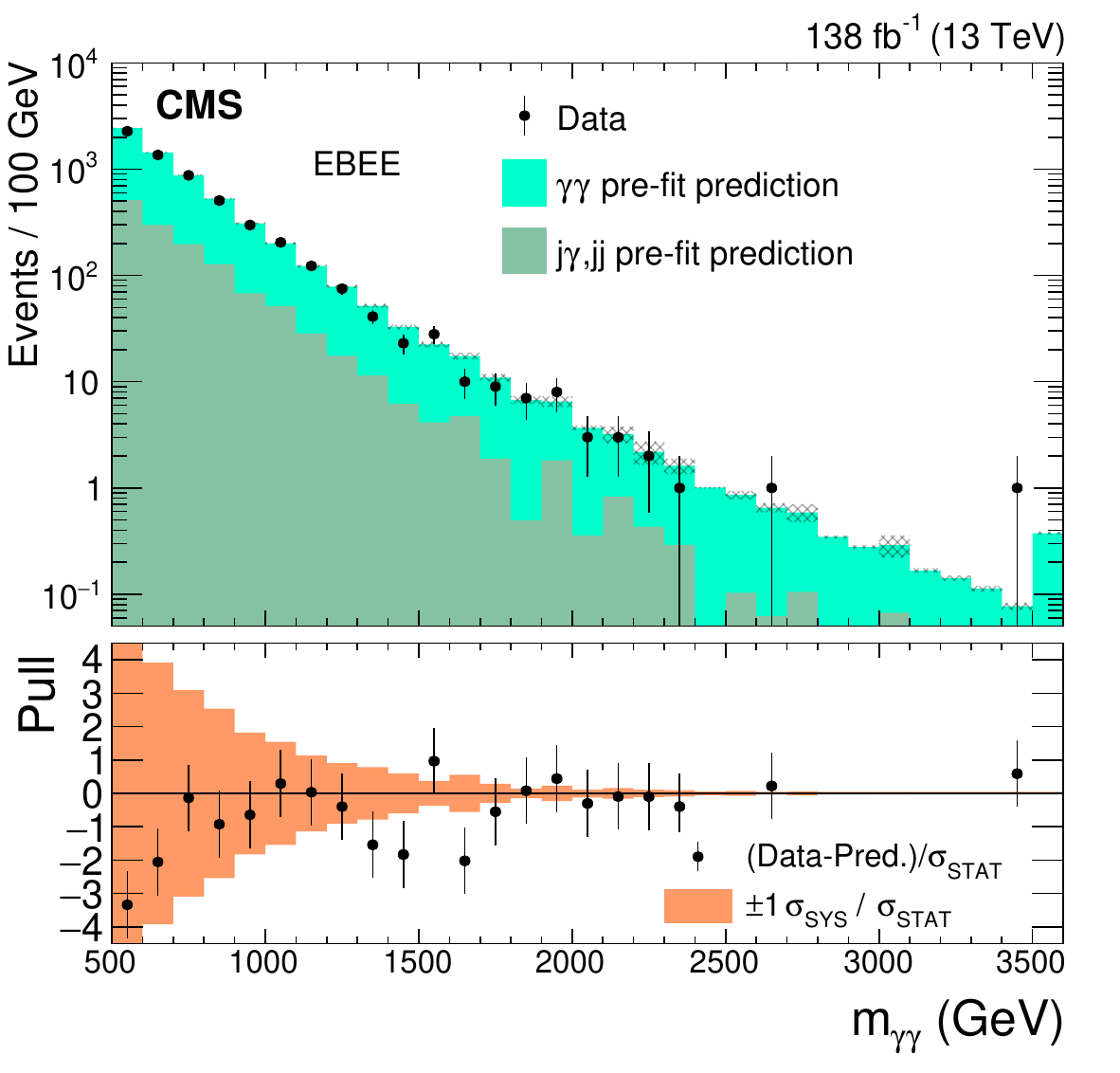}
\caption{The \mgg spectra and the background estimate before nuisance parameter marginalization (``pre-fit'') due to SM diphoton production (\gmgm) and misidentified photon production (j\PGg, jj) for the EBEB (left) and EBEE (right) cases, combining the 2016, 2017, and 2018 data sets.
The pull distributions, defined as the data minus prediction divided by the statistical uncertainty, are shown in the lower panel.
The shaded bands show the systematic uncertainties, neglecting the normalization of the diphoton prediction.
The last bin contains the overflow of events with $\mgg>3.5\TeV$.
\label{fig:Prefit}}
\end{figure}

\begin{figure}[!htbp]
\centering
\includegraphics[width=0.48\textwidth]{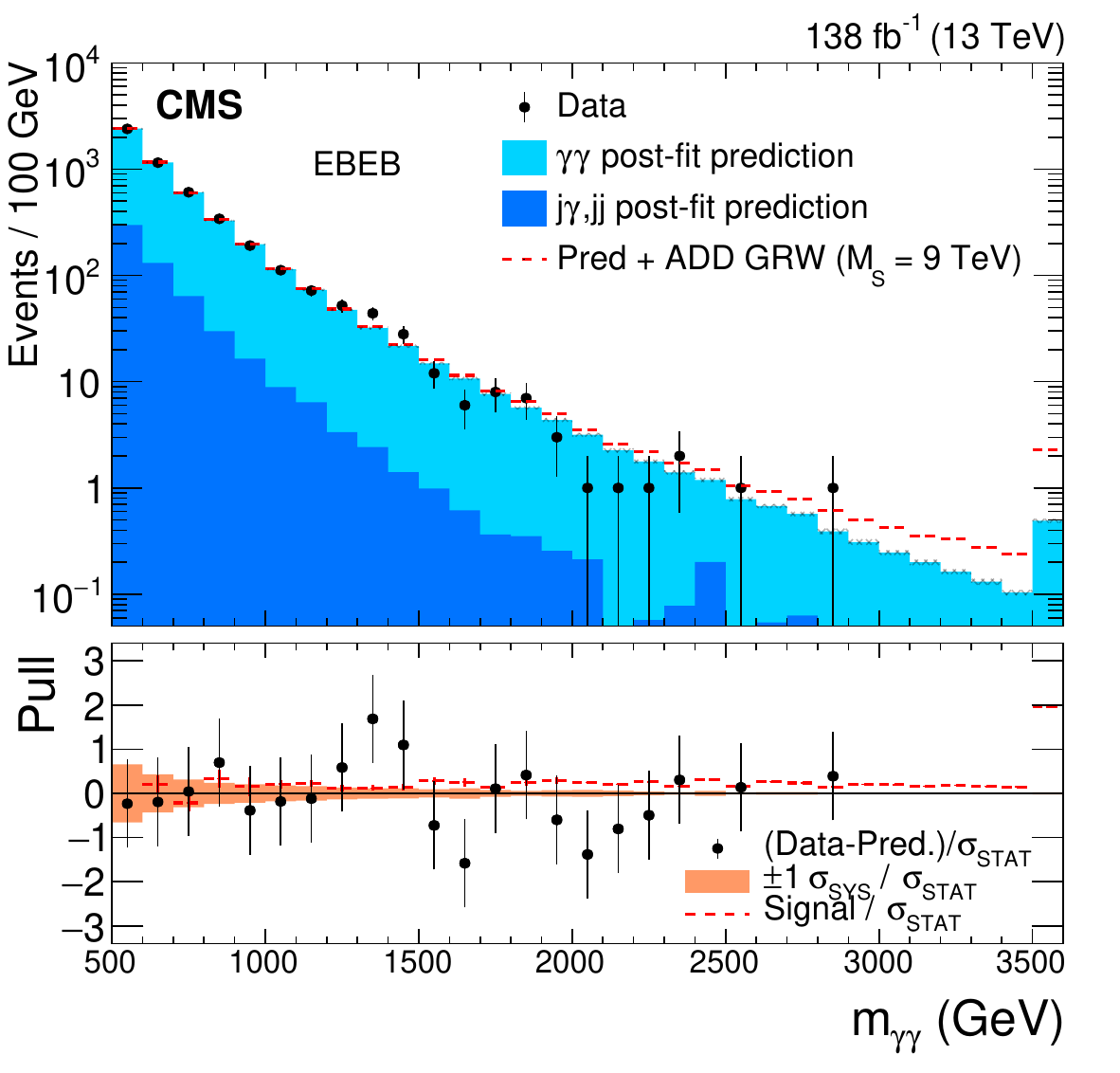}
\includegraphics[width=0.48\textwidth]{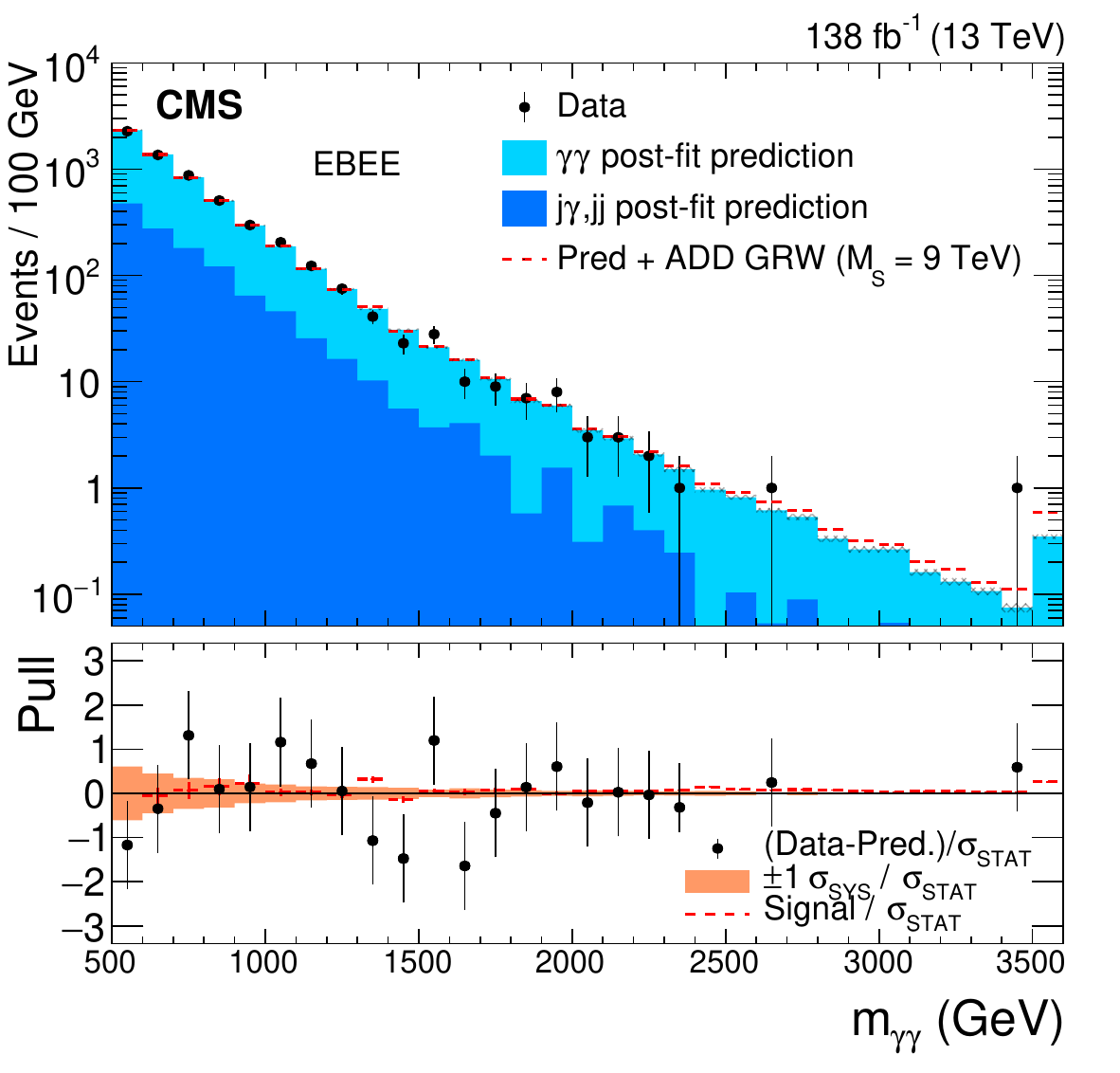}
\caption{The \mgg spectra and background prediction after nuisance parameter marginalization (``post-fit'') due to SM diphoton production (\gmgm) and misidentified photon production (j\PGg, jj) for the EBEB (left) and EBEE (right) cases, combining the 2016, 2017, and 2018 data sets.
The prediction with an ADD signal (GRW convention with $\MS=6\TeV$) is also shown.
The pull distributions, defined as the data minus prediction divided by the statistical uncertainty, are shown in the lower panel.
The shaded bands show the systematic uncertainties, neglecting the normalization of the diphoton prediction.
The last bin contains the overflow of events with $\mgg>3.5\TeV$.
\label{fig:Postfit}}
\end{figure}

Figure~\ref{fig:Prefit} presents the data and background prediction before the marginalization of the nuisance parameters, combined for each of the three years.
The shaded bands show the systematic uncertainties, neglecting the normalization of the diphoton prediction.
Both the previous analysis~\cite{Sirunyan:2018wnk} and the pre-fit predictions in the low-\mgg, background-dominated region give evidence that the NNLO predicts the overall normalization well.
The data and background prediction after the nuisance parameters have been marginalized are shown in Fig.~\ref{fig:Postfit}.

For the ADD model, upper limits are set at the 95\% \CL on the signal strength, which are translated into lower limits on the mass scale \MS.
Table~\ref{Tab:Limits161718} summarizes the results for all ADD model conventions probed.
The excluded values of \MS range from 7.1 to 11.1\TeV, depending on the convention.

\begin{table}[!ht]
  \topcaption{The observed and expected lower limits on \MS in \TeV at the 95\% \CL for different theoretical conventions of the ADD extra dimension model.}\label{Tab:Limits161718}
  \centering
  \renewcommand{\arraystretch}{1.3}
  \cmsTable{\begin{tabular}{cc{c}@{\hspace*{5pt}}cc{c}@{\hspace*{5pt}}cccccc}\hline
    Signal: & GRW && \multicolumn{2}{c}{Hewett} && \multicolumn{5}{c}{HLZ} \\
    \cline{2-2}\cline{4-5}\cline{7-11}
     & &&  negative & positive && $\nED=3$ & $\nED=4$ & $\nED=5$ & $\nED=6$ & $\nED=7$ \\ \hline
    Expected: & $8.7^{+0.7}_{-0.6}$ && $7.3^{+0.3}_{-0.3}$ & $7.8^{+0.6}_{-0.5}$ && $10.3^{+0.8}_{-0.7}$ & $8.7^{+0.7}_{-0.6}$ & $7.9^{+0.6}_{-0.5}$ & $7.3^{+0.6}_{-0.5}$ & $6.9^{+0.6}_{-0.5}$ \\
    Observed: & 9.3 && 7.1 & 8.3 && 11.1 & 9.3 & 8.4 & 7.8 & 7.4 \\ \hline
  \end{tabular}}
\end{table}

\begin{figure}[!ht]\centering
\includegraphics[width=0.50\linewidth]{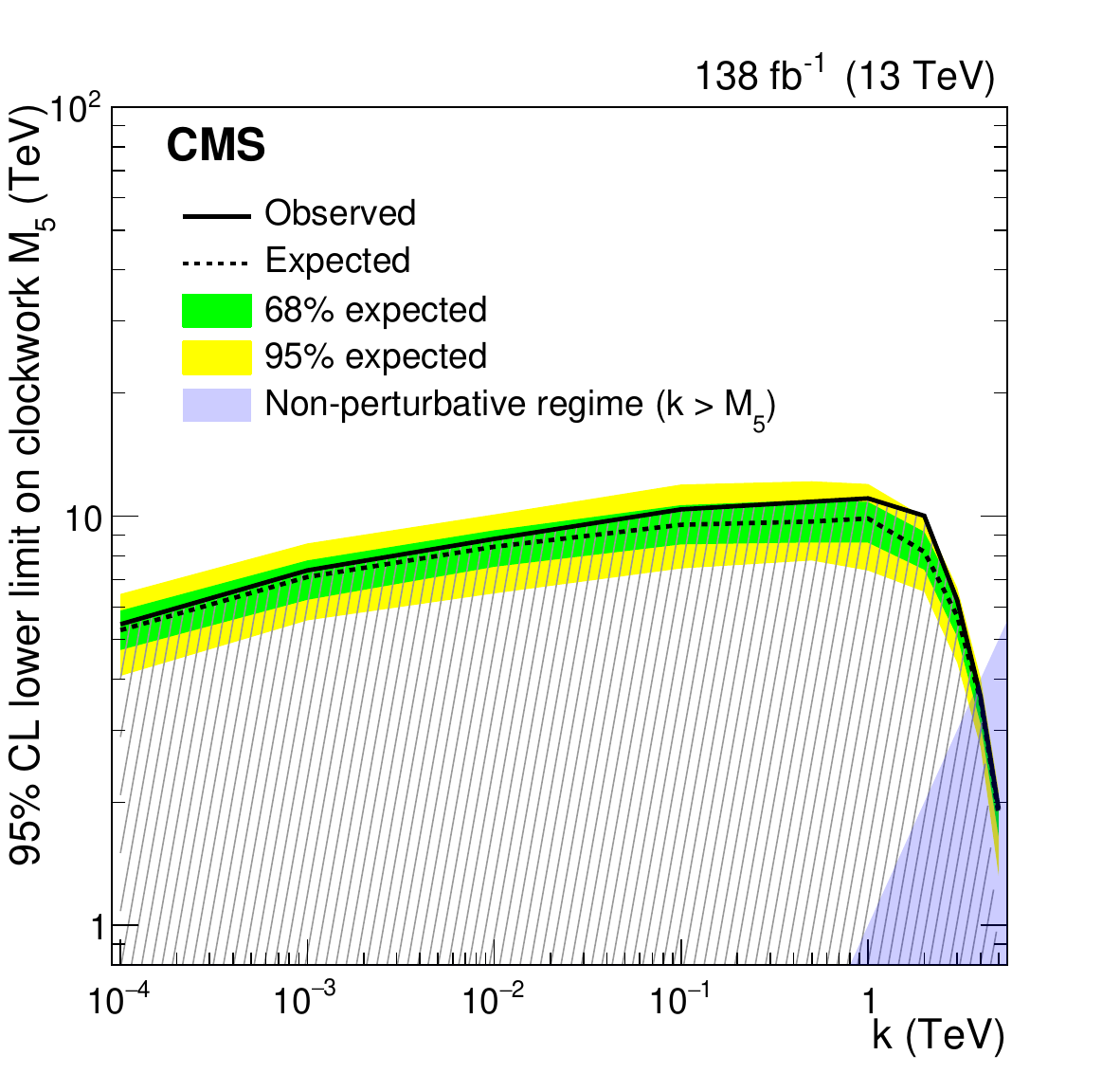}
\caption{The exclusion limit for the clockwork framework over the $k$--\Mfive parameter space.
The darker shaded region denotes where the theory becomes nonperturbative.
The region below the solid line constitutes the excluded region.
Expected 68\% and 95\% limit bands are shown in green and yellow, respectively.}
\label{Fig:LIMIT_Clockwork}
\end{figure}

The limit-setting strategy for the clockwork model is similar to that for the ADD model.
Figure~\ref{Fig:LIMIT_Clockwork} shows the 95\% \CL exclusion limits in the $k$--\Mfive plane.
Values of \Mfive lower than 8.0\TeV are excluded for $k$ values in the range of 0.2\GeV to 2.0\TeV.

As described above, the two different background estimation approaches in this paper are sensitive to different types of deviations from the SM expectation.
Despite the different techniques used, the two methods agree bin-by-bin typically well within the statistical uncertainties, with relative deviations as large as 7\% for $\mgg<1500\GeV$ in both the EBEB and EBEE spectra.
\clearpage

\section{Summary}\label{sec:summary}

A search has been performed for new physics in high-mass diphoton events from proton-proton collisions at a center-of-mass energy of 13\TeV.
The data used correspond to an integrated luminosity of 138\fbinv collected with the CMS detector in 2016--2018.
No statistically significant excess, either resonant or nonresonant, is observed in the spectra.
Masses below 2.2 to 5.6\TeV are excluded at the 95\% confidence level for the excited state of the Randall--Sundrum (RS) graviton, for coupling parameters between $0.01<\widetilde{k}<0.2$.
Limits are also set on the production of scalar Higgs boson like resonances.
In the model with large extra spatial dimensions by Arkani-Hamed, Dimopoulos, and Dvali (ADD), exclusion limits on the mass scale \MS range between 7.1 to 11.1\TeV, depending on the specific convention.
Additionally, exclusion limits are set in the two-dimensional space of the continuum clockwork model, with the fundamental scale \Mfive excluded at the 95\% confidence level below 8.0\TeV for $k$ values between 0.2\GeV and 2.0\TeV.
The present results are the strongest limits to date on ADD extra dimensions and RS gravitons with $\widetilde{k}\ge 0.1$.

\begin{acknowledgments}
  We congratulate our colleagues in the CERN accelerator departments for the excellent performance of the LHC and thank the technical and administrative staffs at CERN and at other CMS institutes for their contributions to the success of the CMS effort. In addition, we gratefully acknowledge the computing centers and personnel of the Worldwide LHC Computing Grid and other centers for delivering so effectively the computing infrastructure essential to our analyses. Finally, we acknowledge the enduring support for the construction and operation of the LHC, the CMS detector, and the supporting computing infrastructure provided by the following funding agencies: SC (Armenia), BMBWF and FWF (Austria); FNRS and FWO (Belgium); CNPq, CAPES, FAPERJ, FAPERGS, and FAPESP (Brazil); MES and BNSF (Bulgaria); CERN; CAS, MoST, and NSFC (China); MINCIENCIAS (Colombia); MSES and CSF (Croatia); RIF (Cyprus); SENESCYT (Ecuador); ERC PRG, RVTT3 and MoER TK202 (Estonia); Academy of Finland, MEC, and HIP (Finland); CEA and CNRS/IN2P3 (France); SRNSF (Georgia); BMBF, DFG, and HGF (Germany); GSRI (Greece); NKFIH (Hungary); DAE and DST (India); IPM (Iran); SFI (Ireland); INFN (Italy); MSIP and NRF (Republic of Korea); MES (Latvia); LMTLT (Lithuania); MOE and UM (Malaysia); BUAP, CINVESTAV, CONACYT, LNS, SEP, and UASLP-FAI (Mexico); MOS (Montenegro); MBIE (New Zealand); PAEC (Pakistan); MES and NSC (Poland); FCT (Portugal);  MESTD (Serbia); MCIN/AEI and PCTI (Spain); MOSTR (Sri Lanka); Swiss Funding Agencies (Switzerland); MST (Taipei); MHESI and NSTDA (Thailand); TUBITAK and TENMAK (Turkey); NASU (Ukraine); STFC (United Kingdom); DOE and NSF (USA).
  
  \hyphenation{Rachada-pisek} Individuals have received support from the Marie-Curie program and the European Research Council and Horizon 2020 Grant, contract Nos.\ 675440, 724704, 752730, 758316, 765710, 824093, 101115353, 101002207, and COST Action CA16108 (European Union); the Leventis Foundation; the Alfred P.\ Sloan Foundation; the Alexander von Humboldt Foundation; the Science Committee, project no. 22rl-037 (Armenia); the Belgian Federal Science Policy Office; the Fonds pour la Formation \`a la Recherche dans l'Industrie et dans l'Agriculture (FRIA-Belgium); the Agentschap voor Innovatie door Wetenschap en Technologie (IWT-Belgium); the F.R.S.-FNRS and FWO (Belgium) under the ``Excellence of Science -- EOS" -- be.h project n.\ 30820817; the Beijing Municipal Science \& Technology Commission, No. Z191100007219010 and Fundamental Research Funds for the Central Universities (China); the Ministry of Education, Youth and Sports (MEYS) of the Czech Republic; the Shota Rustaveli National Science Foundation, grant FR-22-985 (Georgia); the Deutsche Forschungsgemeinschaft (DFG), under Germany's Excellence Strategy -- EXC 2121 ``Quantum Universe" -- 390833306, and under project number 400140256 - GRK2497; the Hellenic Foundation for Research and Innovation (HFRI), Project Number 2288 (Greece); the Hungarian Academy of Sciences, the New National Excellence Program - \'UNKP, the NKFIH research grants K 131991, K 133046, K 138136, K 143460, K 143477, K 146913, K 146914, K 147048, 2020-2.2.1-ED-2021-00181, and TKP2021-NKTA-64 (Hungary); the Council of Science and Industrial Research, India; ICSC -- National Research Center for High Performance Computing, Big Data and Quantum Computing and FAIR -- Future Artificial Intelligence Research, funded by the NextGenerationEU program (Italy); the Latvian Council of Science; the Ministry of Education and Science, project no. 2022/WK/14, and the National Science Center, contracts Opus 2021/41/B/ST2/01369 and 2021/43/B/ST2/01552 (Poland); the Funda\c{c}\~ao para a Ci\^encia e a Tecnologia, grant CEECIND/01334/2018 (Portugal); the National Priorities Research Program by Qatar National Research Fund; MCIN/AEI/10.13039/501100011033, ERDF ``a way of making Europe", and the Programa Estatal de Fomento de la Investigaci{\'o}n Cient{\'i}fica y T{\'e}cnica de Excelencia Mar\'{\i}a de Maeztu, grant MDM-2017-0765 and Programa Severo Ochoa del Principado de Asturias (Spain); the Chulalongkorn Academic into Its 2nd Century Project Advancement Project, and the National Science, Research and Innovation Fund via the Program Management Unit for Human Resources \& Institutional Development, Research and Innovation, grant B37G660013 (Thailand); the Kavli Foundation; the Nvidia Corporation; the SuperMicro Corporation; the Welch Foundation, contract C-1845; and the Weston Havens Foundation (USA).
\end{acknowledgments}

\bibliography{auto_generated}
\cleardoublepage \appendix\section{The CMS Collaboration \label{app:collab}}\begin{sloppypar}\hyphenpenalty=5000\widowpenalty=500\clubpenalty=5000
\cmsinstitute{Yerevan Physics Institute, Yerevan, Armenia}
{\tolerance=6000
A.~Hayrapetyan, A.~Tumasyan\cmsAuthorMark{1}\cmsorcid{0009-0000-0684-6742}
\par}
\cmsinstitute{Institut f\"{u}r Hochenergiephysik, Vienna, Austria}
{\tolerance=6000
W.~Adam\cmsorcid{0000-0001-9099-4341}, J.W.~Andrejkovic, T.~Bergauer\cmsorcid{0000-0002-5786-0293}, S.~Chatterjee\cmsorcid{0000-0003-2660-0349}, K.~Damanakis\cmsorcid{0000-0001-5389-2872}, M.~Dragicevic\cmsorcid{0000-0003-1967-6783}, P.S.~Hussain\cmsorcid{0000-0002-4825-5278}, M.~Jeitler\cmsAuthorMark{2}\cmsorcid{0000-0002-5141-9560}, N.~Krammer\cmsorcid{0000-0002-0548-0985}, A.~Li\cmsorcid{0000-0002-4547-116X}, D.~Liko\cmsorcid{0000-0002-3380-473X}, I.~Mikulec\cmsorcid{0000-0003-0385-2746}, J.~Schieck\cmsAuthorMark{2}\cmsorcid{0000-0002-1058-8093}, R.~Sch\"{o}fbeck\cmsorcid{0000-0002-2332-8784}, D.~Schwarz\cmsorcid{0000-0002-3821-7331}, M.~Sonawane\cmsorcid{0000-0003-0510-7010}, S.~Templ\cmsorcid{0000-0003-3137-5692}, W.~Waltenberger\cmsorcid{0000-0002-6215-7228}, C.-E.~Wulz\cmsAuthorMark{2}\cmsorcid{0000-0001-9226-5812}
% ...
\par}
\vskip\cmsinstskip
\dag:~Deceased\\
$^{1}$Also at Yerevan State University, Yerevan, Armenia\\
$^{2}$Also at TU Wien, Vienna, Austria\\
% ...
\end{sloppypar}
\end{document}